\documentclass[preprint,twocolumn,5p,authoryear,colorlinks=true]{elsarticle}

\journal{Planetary and Space Science}
\usepackage{numcompress}

\usepackage{lineno,hyperref,siunitx,mhchem,booktabs,cuted}
\modulolinenumbers[1]

\usepackage[labelfont=bf,justification=justified,singlelinecheck=false]{caption}
\makeatletter
\g@addto@macro\@floatboxreset\centering
\makeatother

\newcommand{\cbox}[1]{\textcircled{#1}}
\newcommand{\doi}[1]{%
  \textit{DOI:} %
    \href{https://dx.doi.org/#1%
      }{#1%
    }%
}

\begin{document}

\begin{frontmatter}
\title{Transparency of \SI{2}{\um} window of Titan's atmosphere}

\author[GSMA]{Pascal Rannou\corref{correspondingauthor}}\ead{pascal.rannou@univ-reims.fr}
\author[GSMA]{Beno\^{i}t Seignovert}
\author[LPGN]{S. Le Mou\'{e}lic}
\author[Nature]{L. Maltagliati}
\author[GSMA]{M. Rey}
\author[JPL]{C. Sotin}

\address[GSMA]{GSMA, Universit\'{e} de Reims Champagne-Ardenne, UMR 7331-GSMA, 51687 Reims, France}
\address[LPGN]{LPGN, UMR CNRS 6112, Universit\'e de Nantes, Nantes, France}
\address[Nature]{Nature Publishing Group, London, United-Kingdom}
\address[JPL]{Jet Propulsion Laboratory, California Institute of Technology, Pasadena, CA 91109, USA}

\cortext[correspondingauthor]{Corresponding author}

\begin{abstract}
Titan's atmosphere is optically thick and hides the surface and the lower layers from the view at almost all wavelengths. However, because gaseous absorptions are spectrally selective, some narrow spectral intervals are relatively transparent and allow to probe the surface.
To use these intervals (called windows) a good knowledge of atmospheric absorption is necessary. Once gas spectroscopic linelists are well established, the absorption inside windows depends on the way the far wings of the methane absorption lines are cut-off. We know that the intensity in all the windows can be explained with the same cut-off parameters, except for the window at \SI{2}{\um}.
This discrepancy is generally treated with a workaround which consists in using a different cut-off description for this specific window. This window is relatively transparent and surface may have specific spectral signatures that could be detected. Thus, a good knowledge of atmosphere opacities is essential and our scope is to better understand what causes the difference between the \SI{2}{\um} window and the other windows.
In this work, we used scattered light at the limb and transmissions in occultation observed with VIMS (Visual Infrared Mapping Spectrometer) onboard Cassini, around the \SI{2}{\um} window. Data shows an absorption feature that participates to the shape of this window.
Our atmospheric model fits well the VIMS data at \SI{2}{\um} with the same cut-off than for the other windows, provided an additional absorption is introduced in the middle of the window around \SI{\sim 2.065}{\um}.  It explains well the discrepancy between the cut-off used at \SI{2}{\um}, and we show that a gas with a fairly constant mixing ratio, possibly ethane, may be the cause of this absorption. Finally, we studied the impact of this absorption on the retrieval of the surface reflectivity and found that it is significant.
\end{abstract}

\begin{keyword}
Titan \sep Atmosphere

\doi{10.1016/j.pss.2017.11.015}
\end{keyword}

\end{frontmatter}


\section{Introduction}
Titan, the largest satellite of Saturn, has a dense atmosphere of \SI{1.44}{\bar} and a thick haze layer of photochemical aerosols which hides the lower layers and the surface from view. Before the arrival of Cassini, the surface could be probed in near infrared by HST \citep{Smith1996} and with ground based telescope thanks to the progress of adaptive optics \citep[e.g., ][]{Combes1997, Coustenis2001, Hirtzig2005}. Spectroscopic observations and spectro-imaging observations were used to retrieve surface reflectivity \citep{Griffith1991, Griffith2003, Coustenis1995, Negrao2006,Negrao2007} with the information available at that time concerning methane absorption. During the same period of time, clouds were observed first with telescope spectrosopic observations \citep{Griffith1998,Griffith2000} and then imaged \citep{Brown2002, Schaller2005, Hirtzig2009}. In some cases, radiative transfer model were use to determine cloud properties.

Cassini observations allowed us to obtain spectroscopic observations on a broad spectral range, with a good spatial resolution. With the information collected by Huygens during its descent and thanks to the advance in the knowledge of methane spectroscopic data, it is now possible to much  better constrain surface properties inside the methane window \citep[e.g., ][]{Griffith2012, Hirtzig2013} and to retrieve cloud characteristics as drop size, cloud altitude and opacity \citep{Griffith2005,Griffith2006,LeMouelic2012}.

Concerning Titan, one important issue came out from these studies. As for all planetary cases, the spectroscopic linelists which describes the gas absorption is composed of lines which can be represented as Dirac functions. Each line must be convolved by a specific widening profile to account for collisional and Doppler widening. For atmospheric purpose, when data is given with a moderate spectral resolution as in this study, a Voigt profile can be used.
Far from the center of the line, this function turns to a slowly decreasing Lorentzian function which must be modified to decrease faster. To do so, a cut-off function is applied given limit of $|\nu-\nu_0|$, as for instance an exponential decay ($\exp(-|\nu-\nu_0|/\gamma)$) or any other decreasing function, to the Voigt profile \citep[e.g., ][]{deBergh2012}. Cut-off can be more sophisticated as for instance with two limits and two different decays \citep{Hartmann2002} or different cut-off functions.

However, it appeared from studies of Titan photometry that the cut-off parameters which must be applied to the line profiles are different at \SI{2}{\um} window than at other wavelengths \citep{Griffith2012, Hirtzig2013}. This is especially apparent when probing thick clouds because, in this case, when the atmosphere is not correctly set, surface reflectivity can no longer compensate the errors to get the correct spectral shape of the window.
So far, to overcome this difficulty, several works were performed assuming a special wing cut-off at \SI{2}{\um}, different from the wing cut-off in all other windows. This is not satisfactory because no clear reasons are given to justify such a difference and, moreover, for other planets such an adaptation from window to window is not necessary \citep[e.g., ][]{Sromovsky2012, Fedorova2015}.

Defining clearly the opacity inside the windows is especially important to retrieve accurately the surface albedo. We can either retrieve a single value for each window or a small part of spectra inside wide windows (that is beyond \SI{1.5}{\um}). Conclusions on the presence of water ice, the strength of its signatures, or signatures of other components may be drawn from analysis of small parts of surface spectra collected in near infrared windows \citep[e.g., ][]{Griffith2003,Hirtzig2013,Solomonidou2016}.
The retrieved surface albedo and eventually its spectral shape is strongly dependent on the atmosphere opacities, including gas opacity and thus on the cut-off prescription \citep{Negrao2006, Negrao2007, Solomonidou2014}. Although albedo inside windows have not been discussed yet in literature, probably because atmosphere opacity remains too uncertain, it worths mentioning that they can hold valuable information provided that the opacity of the atmosphere above is correctly set.

In this work, we specifically study the intensity of scattered light at the limb of Titan, in the stratosphere in the \SI{2}{\um} window. There, the intensity essentially depends on the aerosol haze and gas properties. Doing so, we avoid the influence of the surface reflectivity, which remains essentially unknown. Our goal is to determine the components which produce the shape of the \SI{2}{\um} window and to define the conditions that would allow to model this window with the same cut-off parameters than for the other windows.

\section{Data and model}
\subsection{Description of the data}
\begin{figure*}[!ht]
  \includegraphics[width=.85\textwidth]{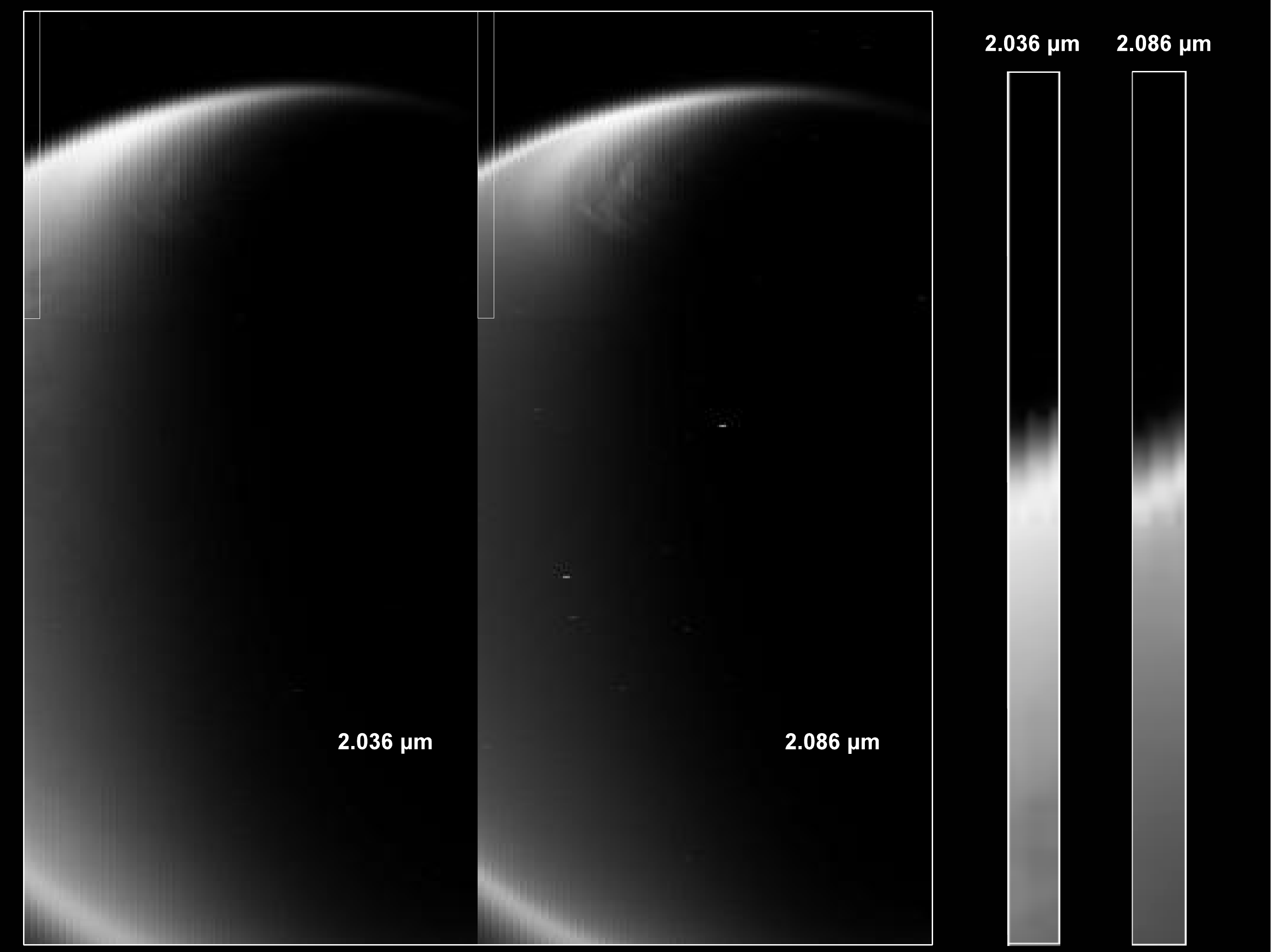}
  \caption{Mosaic of VIMS images of Titan taken inside the methane \SI{2}{\um} window during the flyby T22, at two wavelengths (channels \#167 and \#170). For each wavelength, we used the pixels located in the three columns at the left hand side of the image, shown by the elongated rectangle. At the right side are shown a close up of the selected pixels at the same wavelengths.}
  \label{Images_T22}
\end{figure*}

The data analyzed in this paper is a mosaic of cubes from V1545974724\_1 to V1545983419\_1 taken by  VIMS  during the T22 flyby the 22 December 2006. This image shows the polar region, a large polar cloud, and the atmosphere at the limb of Titan, with a view taken at \ang{113} phase angle \citep{LeMouelic2012}. Thanks to this specific acquisition mode, this image has a very high spatial resolution and it is the only VIMS image that enables us to extract detailed vertical profiles of scattered light at the limb.
The spatial resolution of this image is about \SI{20}{km}, that is half an atmosphere scale height. The data have been calibrated in $I/F$ using the pipeline described in \cite{Barnes2007} and also labelled \emph{RC17} in the description given in \cite{Clark2016}\footnote{\href{http://atmos.nmsu.edu/data\_and\_services/atmospheres\_data/Cassini/logs/vims-radiometric-calibration-pds-2016-v1.20.pdf}{atmos.nmsu.edu/data\_and\_services/atmospheres\_data/Cassini/ logs/vims-radiometric-calibration-pds-2016-v1.20.pdf}}.
We consider that the shift in VIMS channel wavelengths, estimated to be on the order of \SI{2}{nm} at the time of the T22 observation, is negligible for our study.
The \textit{noodle} mode was used for this T22 observation. We reconstructed the corresponding 2D image (Figure~\ref{Images_T22}) by concatenating the whole series of 395 individual $64 \times 1$ pixels cubes. The spectral interval probed by VIMS range from $\lambda=0.3$ to \SI{5.1}{\um}, with 352 channels.
In the infrared part, observed with 256 channels from $\lambda=0.88$ to \SI{5.1}{\um}, the full width at half maximum of the channels is between 0.013 to \SI{0.022}{\um}, thus the spectral resolution is between 120 and 180. For this limb observation, we ignore the wavelength shift of \SI{2.5}{nm} which was observed in the recording of data. In this work, we focus our attention on the radiance factor at the limb in the northern hemisphere.

Although the image seems to probe all the northern polar region, the latitude range at the limb is quite moderate, from \ang{38}N to \ang{55}N at the far edge of the light crescent (Figure~\ref{Images_T22}).
We then extract only one vertical profile as further as possible from the end of the crescent. This set of data is representative of this region of Titan. The part of limb in the southern hemisphere, in the left and lower corner of the image, only extends up to \SI{195}{km}.
This is not enough to perform a study. First, the properties upper part of the atmosphere participate to the scattered intensity at lower altitude. This information is missing in the data.
Secondly, between \SI{195}{km} and the lower level that can be studied with this technique, we would not be able to collect more than 3 of 4 points. This is not enough to study the scattering at the limb.

\begin{figure}[!ht]
  \includegraphics[width=.45\textwidth]{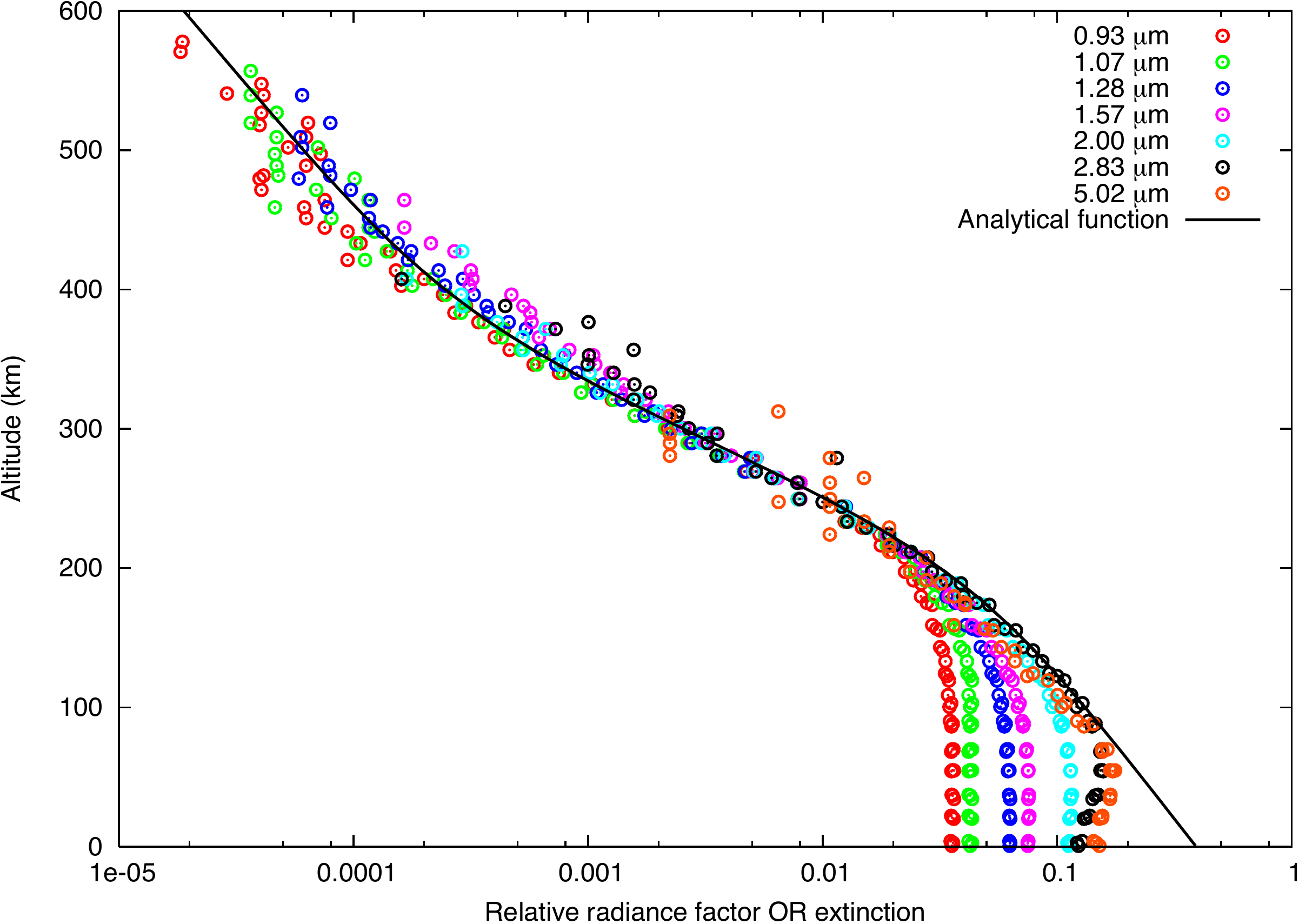}
  \caption{Profiles of radiance factors observed inside the seven windows probed by VIMS-IR. Each point corresponds to the radiance factor of a pixel in the selected zone shown in the Figure~\ref{Images_T22}.
  The continuous line shows the extinction profile from our model that best reproduces the radiance factor profiles.
  All profiles are scaled in this plot to obtain overlapping profiles. This comparison of scaled profiles makes visually appearing the vertical structure of the extinction profile down to \SI{100}{km}, the altitude below which the profile at the most transparent wavelength (\SI{5.03}{\um}) becomes saturated.}
  \label{three_scale_heights}
\end{figure}

For each of the 256 VIMS channels, we consider radiance factors $I/F$ by selecting pixels in the three columns at the left edge of the images. We then obtain vertical $I/F$ profiles, around \ang{40}N, with a spatial resolution of about \SI{20}{km} (Figure~\ref{three_scale_heights}).
A typical $I/F$ profile roughly follows a scale heigh from the top of the atmosphere down to a given level, corresponding to a critical value of the tangential optical depth.
In this part of the profile, the total intensity depends on the integrated opacity along the line of sight and on the average properties of the atmosphere (single scattering albedo, phase function). The multiple scattering may also play a role in the final result.
Beyond the critical tangential opacity, the profile becomes constant or decreases, depending on the actual geometry of the observation, and there, the data does not yield any more information about the atmosphere along all the line of sight.
To study the atmosphere, we essentially consider the part of the profile which follows the scale height because it contains information about the vertical structure. Since haze opacity decreases with wavelength, VIMS is able to probe altitudes up to \SI{500}{km} at \SI{0.88}{\um} and down to \SI{50}{km} at \SI{5}{\um}.
We therefore determine that the haze profile can be described with three different scale heights with transition around \SI{225}{km} and \SI{350}{km} (Figure~\ref{three_scale_heights}).

On Titan, the observed radiance factors bear the spectral signature of methane which produces several alternate bands and windows. The outgoing intensity depends on the methane absorption and on the stratospheric haze. However, while the methane is the main source of opacity in the bands, it is less of a factor for the opacity in windows. As a consequence, the spectral shape of the radiance factor in windows depends on the gaseous continuum. The balance between the haze and methane extinction will be an important issue in our study of the \SI{2}{\um} windows.

Another important issue concerns the absorption by ethane which leaves a significant signature, as demonstrated with occultation observations by \cite{Maltagliati2015}. Several unexplained absorption features appear well correlated to the ethane cross-section \citep{Sharpe2004} measured at ambient temperature and standard pressure. However, it is not possible to perform a valuable analysis because these cross-sections are not suitable for use at all pressures or at the temperature range of Titan. No spectroscopic linelist yet exist in the wavelength range studied in this work. Therefore, to perform our analysis, we decided to remove the spectral intervals where ethane obviously participates to the absorption \citep{Maltagliati2015}, except in the core of the \SI{2}{\um} window which is the scope of this study.

\subsection{Description of the model}

To perform this work, we used the model of scattering at the limb of a planet as described first in \cite{RagesPollack1983}, and used by \cite{Rannou1997,Rannou2006}. In this model, the calculation first consists in summing the single scattered intensity along a line of sight at the limb of the planet assuming a spherical geometry. This step essentially turns into computing the optical thickness and the Beer-Lambert attenuation along the incoming path of the solar photons and the outgoing path of the scattered photons to the probe. For this, we need to discretize the atmosphere in $n$ layers, bounded by $n+1$ levels. The atmosphere properties are supposed to be uniform inside a given layer. In this model, we used 70 layers of \SI{10}{km} thickness.

\begin{figure*}[!ht]
  \includegraphics[width=\textwidth]{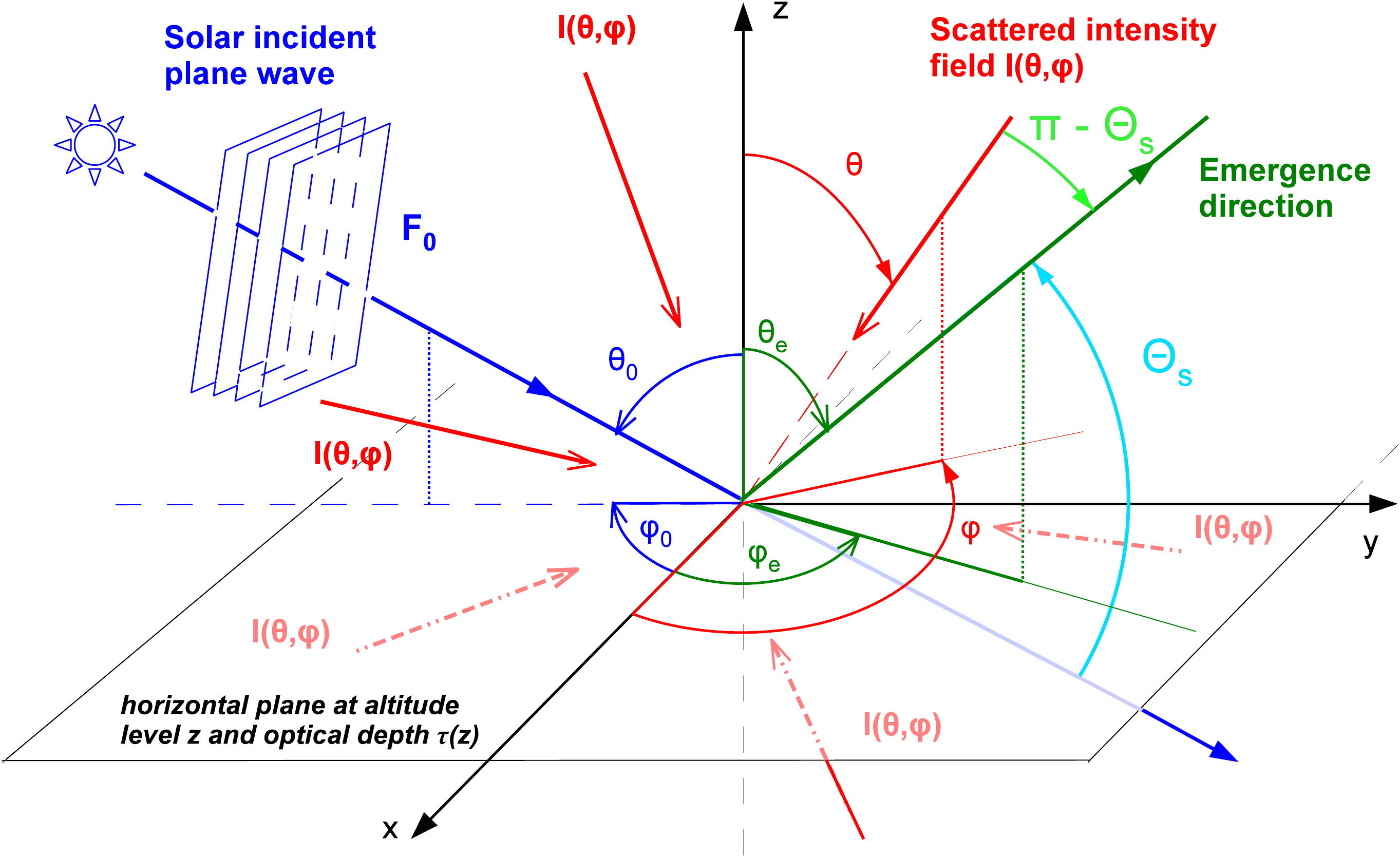}
  \caption{Sketch displaying the geometry of the scattering at a given layer of the atmosphere for the direction of the incoming solar flux $F_0(\tau)$ ($\theta_0$ and $\phi_0$) (blue lines), the scattered intensity field $I(\tau,\theta,\phi)$ (red lines) and the emerging direction $\theta_e$ et $\phi_e$ (green).
The scattering angle $\Theta_s$ between the solar incidence direction and the emergence direction ($\Theta_s(\theta_0,\phi_0,\theta_e,\phi_e)$) is shown in cyan and $\pi - \Theta_s(\theta,\phi,\theta_e,\phi_e)$, the supplementary angle (used for the clarity of the graph) of the scattering angle between a beam of the scattered field and the emergence direction is shown in light green.}
  \label{orientation_rms}
\end{figure*}

In a second step, we evaluate the multiple scattering by using a model of atmosphere, whose properties are representative of Titan. This model evaluates at each level of the atmosphere the total amount of scattered photons toward the observer relatively to the direct photons from the sun scattered once toward the observer. This computation is performed with the relevant geometry for the incident flux and for the outgoing direction at the plane of the limb. We then obtain the multiple to single scattering ratio, $\rho_{ms}(z)$, as a function of the altitude ($z$), defined as:

\begin{strip}
\begin{equation}
\rho_{ms}(z)=\frac{\int\limits_{0}^{\pi}\int\limits_{0}^{2\pi} I(\tau,\theta,\phi)\times
P(\tau,\Theta_{s}[\theta_e,\phi_e,\theta  ,\phi]) d\Omega + F_0 \times \exp(-\tau/\mu_0) \times
P(\tau,\Theta_{s}[\theta_e,\phi_e,\theta_0,\phi_0])}        {F_0 \times \exp(-\tau/\mu_0) \times
P(\tau,\Theta_{s}[\theta_e,\phi_e,\theta_0,\phi_0])}
\end{equation}
\end{strip}

where $\theta$ and $\phi$ indicate a direction of the space relative to the normal direction and relative to an arbitrary azimuth, $\theta_e$ and $\phi_e$ indicates the emergence direction (toward the observer), $\theta_0$ and $\phi_0$ is the incident direction of the solar photons, $\tau$ is the vertical optical depth of the altitude $z$ ( with a value set to 0 at the top of the atmosphere), $P(\tau,\Theta_{s})$ is the phase function at the scattering angle $\Theta_{s}$ and at the level $\tau$. $I(\tau,\theta,\phi)$ is the intensity field at the level $\tau$ and $F_0$ is the solar flux at Titan (Figure~\ref{orientation_rms}).

The integral term is positively-definite or null and it describes the integration of the scattered intensity converging toward one parcel of the atmosphere (this is the scattered part of the actinic flux). The term involving the solar flux $F_0$ is strictly positive and describes the direct solar photons. Thus, we see that in absence of scattered photons ($I(\tau, \theta, \phi) = 0$), $\rho_{ms}(z)$ takes the lowest possible value, that is 1. As soon as there is a contribution of scattered light, $\rho_{ms}(z)$ increases and is boundless. Therefore, multiple scattering always increases the observed intensity relatively to single scattering. A value $\rho_{ms}(z) \gg 1$  indicates that the amount of scattered light largely exceeds the direct light from the Sun and the final intensity that is computed is dominated by multiple scattering.

This estimation, $\rho_{ms}(z)$, is then used as a multiplying factor for the singly scattered light by each volume of atmosphere along the line of sight. The final radiance factor $I/F$ of the $n^{th}$ layer (where $n$=1 is for the upper layer), integrated along a line of sight, is calculated following \cite{RagesPollack1983} with the following equation:

\begin{strip}
\begin{equation}
I/F_n = \sum_{i=1}^{2n} \int_{x_{i-1}}^{x_i} \frac{\varpi_j
P_j(\Theta_{s})}{4} \exp(-\tau^0_i-\tau^1_i)
k(z(x)) \rho_{ms}({z(x))} \mathrm{d}x
\end{equation}
\end{strip}

where the summation is performed on the $2 \times n$ segments which are defined by the intersections of the line of sight and the spherical shells defining the layers boundaries. Each layer of the atmosphere is crossed twice. The impact factor, that is the lowest altitude reached by the line of sight, is given by the bottom of the $n^\mathrm{th}$ layer. $x$ is the abscissa along the line of sight. $\tau^0_i$ and $\tau^1_i$ are the opacities along the incident and emergent path. $\varpi_j$ is the average single scattering of the layer $j$, where the index $j=i$ if $j<n+1$ and $j=2 \times n+1-i$ if $j>n$.

The multiple scattering factor $\rho_{ms}({z(x)})$ is relevant only at the plane of the limb (that is for $x=0$). However, we apply this factor everywhere, using the relevant altitude $z(x)$. To support this approximation, we evaluated that the contribution function along a line of sight is approximately a Gaussian function centered on $x=0$ with a variance $\simeq \sqrt{(R_T+z_0)\times H}$, where $R_T$ is the radius of Titan, $z_0$ is the impact parameter of the observation and $H$ is the haze scale height. This variance is approximately \SI{380}{km}.
We estimate that, in order to account for the change in geometry along the line of sight, one should change the local incident angle by about \ang{\pm 9} around the reference value $\theta_0$ (equal to \ang{23} for this image). This gives a minor effect, first, because the incident angle is close to nadir so the multiple scattering only weakly depends on the incident angle. The value of $\rho_{ms}$ changes by \SI{+0.38}{\percent} per degree and then \SI{\pm 3.4}{\percent} for \ang{\pm 9}.
Secondly, there is an anti-symmetrical effect for the variation of $\rho_{ms}(z)$ around the reference value calculated with positive and negative angular shift which cancels out the difference along the line of sight. We then finally used the value of $\rho_{ms}({z})$ calculated for the geometry at the limb for all the part of the line of sight.

\subsection{Optical properties of aerosols}

To account for haze properties, we followed the description given by \cite{Doose2016}. Hereafter the terms \textit{haze} and \textit{aerosols} are used indifferently to designate the haze layer of photochemical aerosols and the aerosol particles.
The phase functions that we used are therefore modified relatively to those published in \cite{Tomasko2008}. For the other spectral properties, we use a model of scattering by fractal aggregates with optical constants tuned to match the available constraints.
We used the same aggregate characteristics as used by \cite{Doose2016} and \cite{Tomasko2008}, that is aggregates with a fractal dimension $D_f=2$ and with \num{3000} spherical grains (monomers) of \SI{50}{nm}. We then seek the imaginary refractive index that allows us to match the single scattering albedo published by \cite{Doose2016} between \SI{400}{nm} and \SI{900}{nm}.
The single scattering albedo of the photochemical haze above \SI{80}{km} and the mist layer below \SI{80}{km} are significantly different than those published by \cite{Tomasko2008}, and thus we expect to see differences in the refractive indices of the aerosols. We used the same procedure as explained in \cite{Rannou2010}. Once the particle geometry is defined, and assuming the real part of the refractive index from \cite{Khare1984}, the imaginary part of the refractive index $\kappa$ is the only parameter that remains to determine.
Once retrieved, we use the refractive index to compute the extinction cross-sections $\sigma_H$ and the single scattering albedo $\varpi_H$ of the aerosols.

For this study, we leave the vertical structure of the haze as free parameters. $I/F$ profiles at different wavelengths clearly show that three parameters $X_1$, $X_2$ and $X_3$ can be used to described the haze vertical profile above about \SI{100}{km} (Figure~\ref{three_scale_heights}).
Below \SI{80}{km}, we assume a layer of undefined material which has the same properties as defined in \cite{Doose2016} and \cite{Tomasko2008}.
Following \cite{deBergh2012}, we use the same phase function for the mist as for the haze, and we used the same relationship as in \cite{Doose2016} to fix the single scattering albedo of this layer, that is $\varpi_M = (0.565 + \varpi_H)/1.5$ while the spectral slope of the mist is the same as for the haze.

\subsection{Choice of gas absorption linelist}

The gas linelists that we used come from the Hitran database \citep{Rothman2013} for all gases except for methane and its isotopes. The linelists for methane \ce{CH4}, deuterated methane \ce{CH3D} and the \ce{^{13}CH4} are provided by \cite{Rey2013} from theoretical calculations. These linelists have been shown to give good results compared to the empirical band model of \cite{KarkoschkaTomasko2010}, dedicated to match Titan photometry, and do not present any of the flaws that can be observed in Hitran or in Exomol databases \citep{Rothman2013, YurchenkoTennyson2014}. A thorough comparison is performed by  \cite{Rey2017} to evaluate the value of different databases in the frame of Titan photometry.

When a spectroscopic linelist is used for atmospheric purposes, we have to convolve each line by a widening profile. We used a Voigt profile with a cut-off applied at $\Delta \nu_{co} = |\nu-\nu_0|$ from the center of the line $\nu_0$ and a sublorentzian decay $\gamma_{co}$.
We cut-off the Voigt profile by multiplying it by a function $\phi(\Delta \nu=|\nu-\nu_0|)$, where $\phi(\Delta\nu)=1$ if $\Delta\nu < \Delta\nu_{co}$ and $\phi(\Delta\nu)=\phi_0\times \exp(-\Delta\nu/\gamma_{co})$ if $\Delta\nu > \Delta\nu_{co}$ \citep{deBergh2012}.
$\phi_0$ is set in order to ensure the continuity of the function $\phi(\Delta\nu)$. $\Delta \nu_{co}$ and $\gamma_{co}$ are free parameters and they are the main focus of this work. We use as reference values for Titan $\Delta \nu_{co} = \SI{26}{\per\cm}$ and $\gamma_{co} = \SI{120}{\per\cm}$.
This set up gives good match of the intensity at all windows except at \SI{2}{\um}. We use the line broadening parameters for the Lorentzian profile $\alpha_0 = \num{6.5e-2}$ and the exponent $n=0.85$, which are relevant for \ce{CH4} in \ce{N2} \citep{MenardBourcin2007}. All the other rules needed to build coefficients for the gas absorption from the linelists are defined without additional parameters \citep[e.g., ][]{Hanel2003}.
The only other gaseous absorption which participates, marginally, to the outgoing intensity is provided by the pressure induced absorption of \ce{N2-N2} and \ce{N2-H2} dimers at \SI{2}{\um} from \cite{McKellar1989}.

Pressure and temperature profiles needed to compute the gas opacities are those retrieved by the Huygens Atmospheric Structure Instrument (HASI) \citep{Fulchignoni2005} while the methane mixing ratio comes from the latest analysis of the Gas Chromatograph Mass Spectrometer (GCMS) \citep{Niemann2010}.
To treat the gaseous absorption in the radiative transfer model, we used the correlated-k method \citep{Goody1989}, and we used this method with 4 terms for each VIMS spectral channel \citep{Rannou2010}.

\begin{figure*}[!ht]
\includegraphics[width=\textwidth]{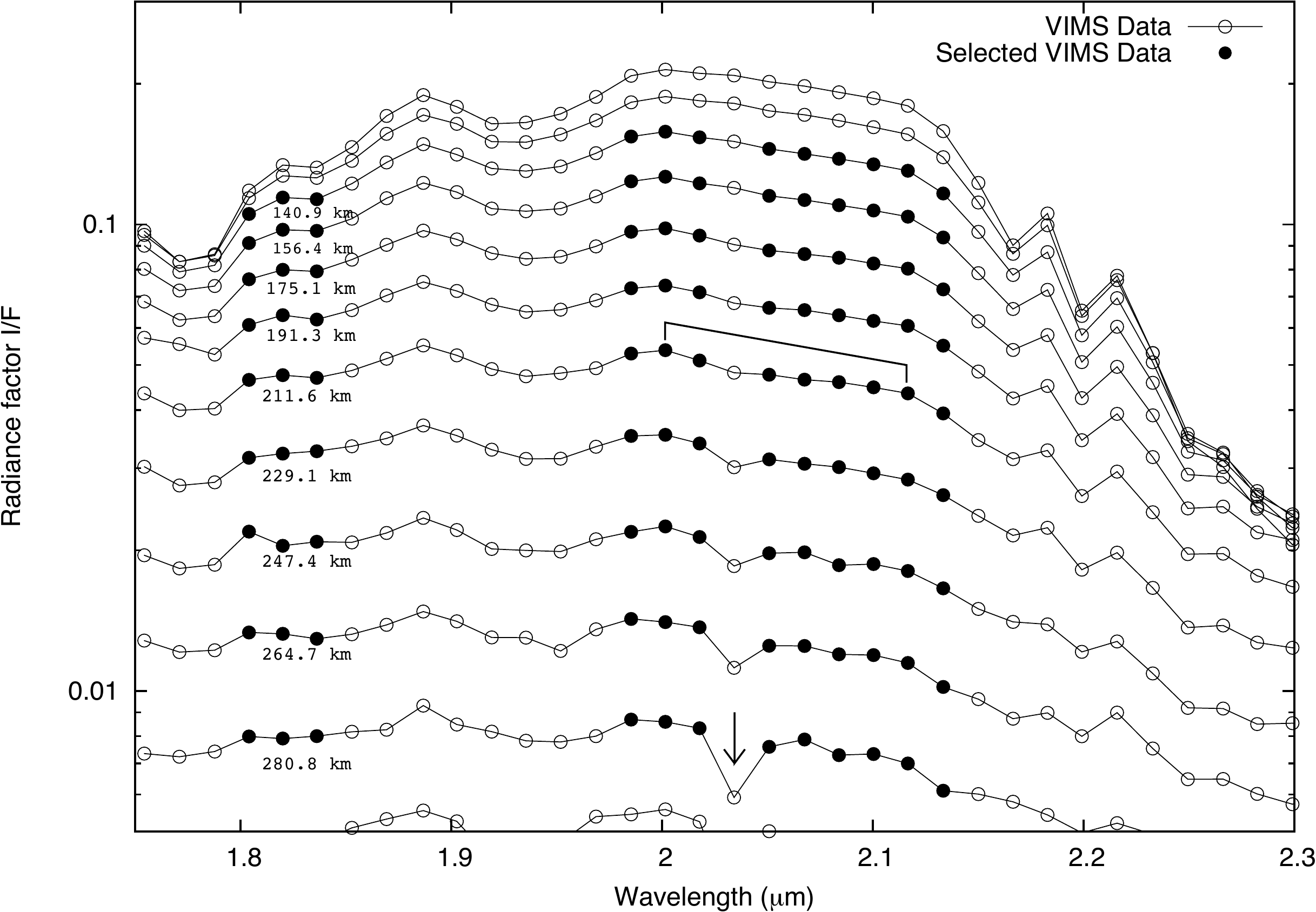}
\caption{ Spectral variation of the radiance factors at several altitudes observed by VIMS. We removed from our analysis the major ethane absorption wavelengths and we remove altitudes where data are too noisy (above \SI{300}{km}) and where the $I/F$ profile becomes saturated (below \SI{130}{km}) (empty circles). Only the data shown with filled circles are actually used in our analysis and for the retrievals.
The spectra we used are labelled with their altitudes, from \SI{140}{km} to \SI{280}{km}. The vertical arrow shows the wavelength corresponding to channel \#167, and the the bracket above the spectra at \SI{211.6}{km} shows the spectral region where an additional absorption is needed, in the middle of the \SI{2}{\um} window. The uncertainty on the radiance factor (\SI{\pm  3}{\percent}) is comparable to the size of the symbols \citep{Sromovsky2012}.}
  \label{spectra_fit0}
\end{figure*}

\section{Study of the \SI{2}{\um} window}

Our computations are restricted to near the \SI{2}{\um} window, between 1.75 and \SI{2.40}{\um}. We exclude four spectral regions ([1.75-\SI{1.79}{\um}], [1.85-\SI{1.97}{\um}], [2.025-\SI{2.045}{\um}] and [2.15-\SI{2.40}{\um}]) from our analysis because they contain undetermined absorption features \citep{Maltagliati2015}.
That leaves us with two intervals for our analysis: [1.79-\SI{1.85}{\um}] and [1.97-\SI{2.15}{\um}], from which we remove the channel \#167. We also restrict the retrieval to the altitude range between 130 and \SI{380}{km}.
The upper and lower limits are defined respectively by the noise level and the level below which the $I/F$ profile do no longer follow a scale height.

\begin{figure}[!ht]
\includegraphics[width=.45\textwidth]{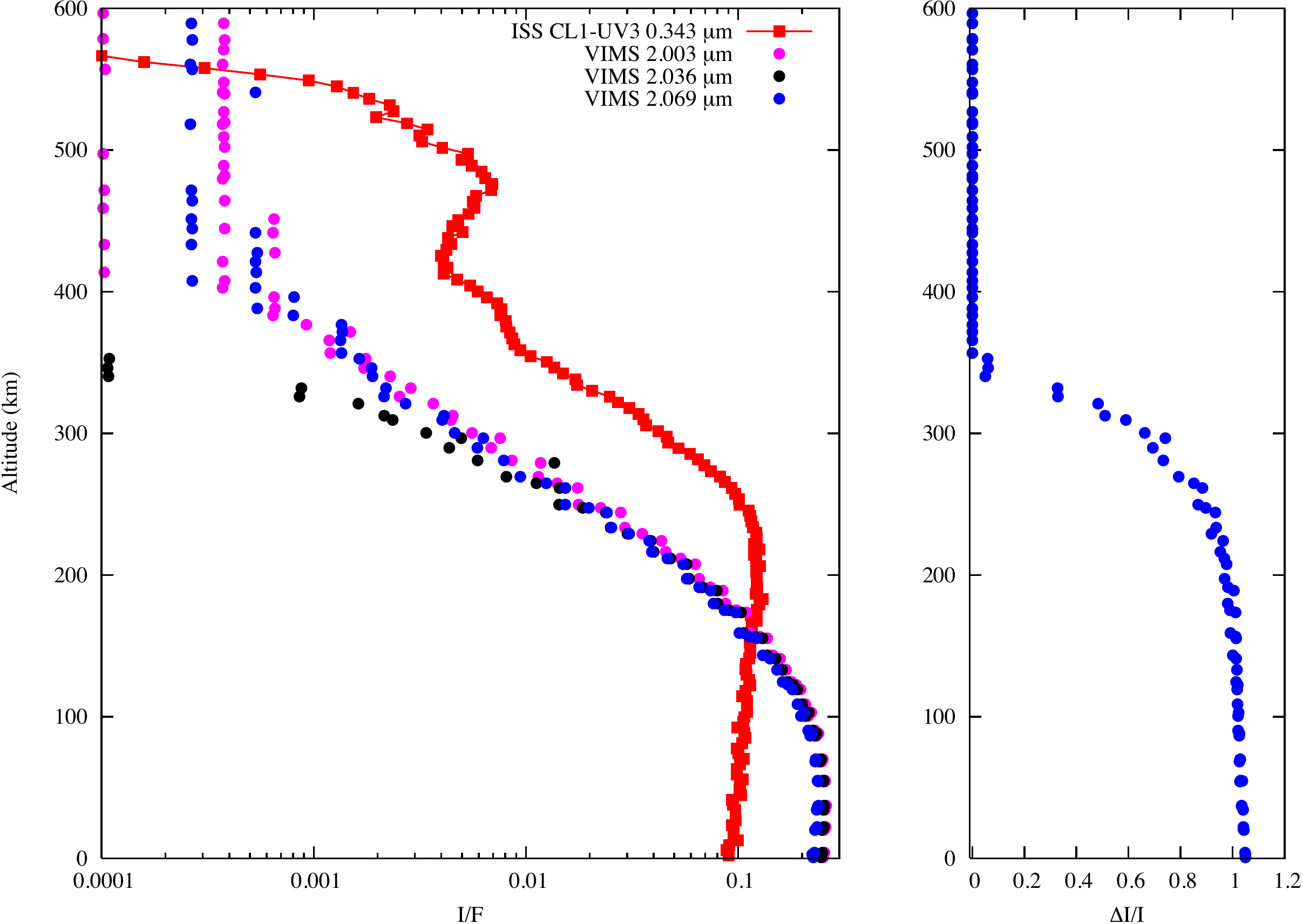}
\caption{At left: Profiles of radiance factor profile at \SI{2.03}{\um} (channel \#167) and at two surrounding wavelengths. The $I/F$ profile decreases faster at \SI{2.03}{\um} than in the two other profiles above \SI{220}{km} thus showing the distribution of the peak.
There is no apparent correlation with the haze distribution, shown with the $I/F$ profile at \SI{0.343}{\um} from the Cassini Imaging and Science Subsystem (ISS).
At right: Relative strength of the peak computed as $\Delta I/I = (I/F_{167} - <I/F>) /<I/F>$ where $<I/F>$ is a average value of $I/F$ over the 6 channels from \#164 to \#170, excepted \#167.}
 \label{profil_peak_203}
 \end{figure}

In the data, we readily see that a feature at \SI{2.03}{\um} (channel \#167), which is very sharp at high altitudes, smooths and widens with decreasing altitudes (Figure~\ref{spectra_fit0}). This feature is located in the middle of the \SI{2}{\um} window and apparently affect the atmosphere opacity as to produce the observed discrepancy in spectroscopic properties \citep{Bailey2011, Griffith2012, Hirtzig2013}.
At lower altitude, we can see that $I/F$ follows two different slopes, with an inflection around \SI{2.03}{\um}. Even if we disregard the channel \#167, this behaviour involves all the channels between 2 and \SI{2.1}{\um}. Therefore, it can hardly be attributed to an instrumental effect. Such structure also appear in other images at much lower spatial resolution and with even more prominent signatures.
Because this is inside a methane window, methane can not produce such an inflexion. Only a supplementary absorption can explain this. It shows that the \SI{2}{\um} window is not smooth and featureless as generally thought and this features may also affect the spectra in close nadir viewing.
Figure~\ref{profil_peak_203} displays the radiance factor obtained at the wavelength of the peak (channel \#167, i.e. \SI{2.036}{\um}) and in two surrounding channels. It shows a marked transition in the peak strength above \SI{250}{km}, and the peak becomes very sharp above \SI{300}{km}. It does not seem directly correlated to the haze structure.
The relative difference of $I/F$ between the channels \#167 and the average of the surrounding channels (from \#163 to \#166 and from \#168 to \#170) quantifies the rapid change in the mesosphere, above \SI{300}{km}. We checked if this peak is real or may be due to a problem in VIMS spectra.
To do so, we sought for this peak elsewhere in the same VIMS image, especially in the terminator side where levels of intensity are similar to the limb. When mapping the peak signature, as for instance the intensity relative differences as plotted in Figure~\ref{profil_peak_203}, we clearly observe discontinuities linked to the composite nature of the image. These discontinuities come from the image in the channel \#167 and are not observed in other channels. This may have an instrumental origin linked to different sequences of observations. Although the peak itself may be real, we do not account for it in our analysis.

\subsection{Shape of the window}
In this first step, our purpose is to use a model where haze and gases are treated in an usual manner, as described previously, in order to study the impact of line profile cut-off characteristics (i.e., $\Delta \nu_{co}$ and $\gamma_{co}$).
Here the shape of the window is only a consequence of the haze properties and the gas extinction. In windows, the latter is controlled by the cut-off values. In this study, we also have to characterize the vertical profile of the haze extinction.
There are three parameters for the vertical profile, $X_1$, $X_2$, $X_3$, for the altitude regions below \SI{225}{km}, between 225 and \SI{350}{km} and above \SI{350}{km} (Figure~\ref{three_scale_heights}) and a fourth parameters to scale the amount of haze $F_H$.
We note here that the parameters $X_i$ do not strictly correspond to the scale height except for the two limiting behaviours ($z \rightarrow 0$ and $z \rightarrow \infty$). We define an analytical function to model the haze vertical profile with the following expression for the extinction:

\begin{strip}
\begin{equation}\label{eq:extinction}
\begin{split}
k(z,\lambda) = A\ F_H\ \tau_{ref}(\lambda_0)\ \frac{\sigma_{ext}(\lambda)}{\sigma_{ext}(\lambda_0)}\times
\Bigg[ \left(\frac{1}{\exp(-(z-z_1)/X_1)}+\frac{1}{\exp(-(z-z_1)/X_2)}\right)^{-1}\\
+ \left(\frac{1}{\exp(-(z_2-z_1)/X_1)}+\frac{1}{\exp(-(z_2-z_1)/X_2)}\right)^{-1} \times \exp(-(z-z_2)/X_3) \Bigg]
\end{split}
\end{equation}
\end{strip}

where $z_1=\SI{225}{km}$ and $z_2=\SI{350}{km}$, $\tau_{ref}(\lambda_0)$ is the reference column opacity at $\lambda_0=1\ \mu$m taken from \cite{Doose2016}, $\sigma_{ext}(\lambda)$ is the aerosol cross-section as a function of the wavelength used for this work and $A$ is the normalization factor of the expression between the brackets, integrated up $z=\SI{55}{km}$ to space.
With this definition, $F_H$ is simply a scaling factor for the column opacity of haze relatively to the value published by \cite{Doose2016}, but the vertical distribution is controlled by the analytical function involving the parameters $X_i$ and the spectral variation is given by the model of scattering by fractal aggregates and the new optical constants, as defined previously.
Below $z=\SI{55}{km}$, the layer has a constant extinction and is normalized with the prescription from \cite{Doose2016} as well. Two supplementary parameters concern the methane absorption (the cut-off parameters $\Delta \nu_{co}$ and $\gamma_{co}$). The other properties related to haze (cross-sections, phase function) and gaseous absorption are defined in the previous section.
We then define the main structure of the haze profile by seeking, for each set of $\Delta \nu_{co}$ and $\gamma_{co}$, the set of parameters $X_1$, $X_2$, $X_3$ and $F_H$ which best fit the $I/F$ spectra. We use a Levenberg-Marquardt routine to minimize the $\chi^2$, and then we are also able to evaluate the quality of the solution. The shape of the analytical vertical profile of the extinction (Eq.~\ref{eq:extinction}) is not able to follow accurately all the small scale structures that may be in the real profile.
Thus, to compute the $\chi^2$, we allow a scaling of the model intensity by less than \SI{5}{\percent} to minimize the differences between the model and the data at each level. Doing that, we assume that the real extinction profile may indeed have oscillations around the ideal guess which is displayed in Equation~\ref{eq:extinction}, and account for them.

\begin{figure}[!ht]
  \includegraphics[width=.45\textwidth]{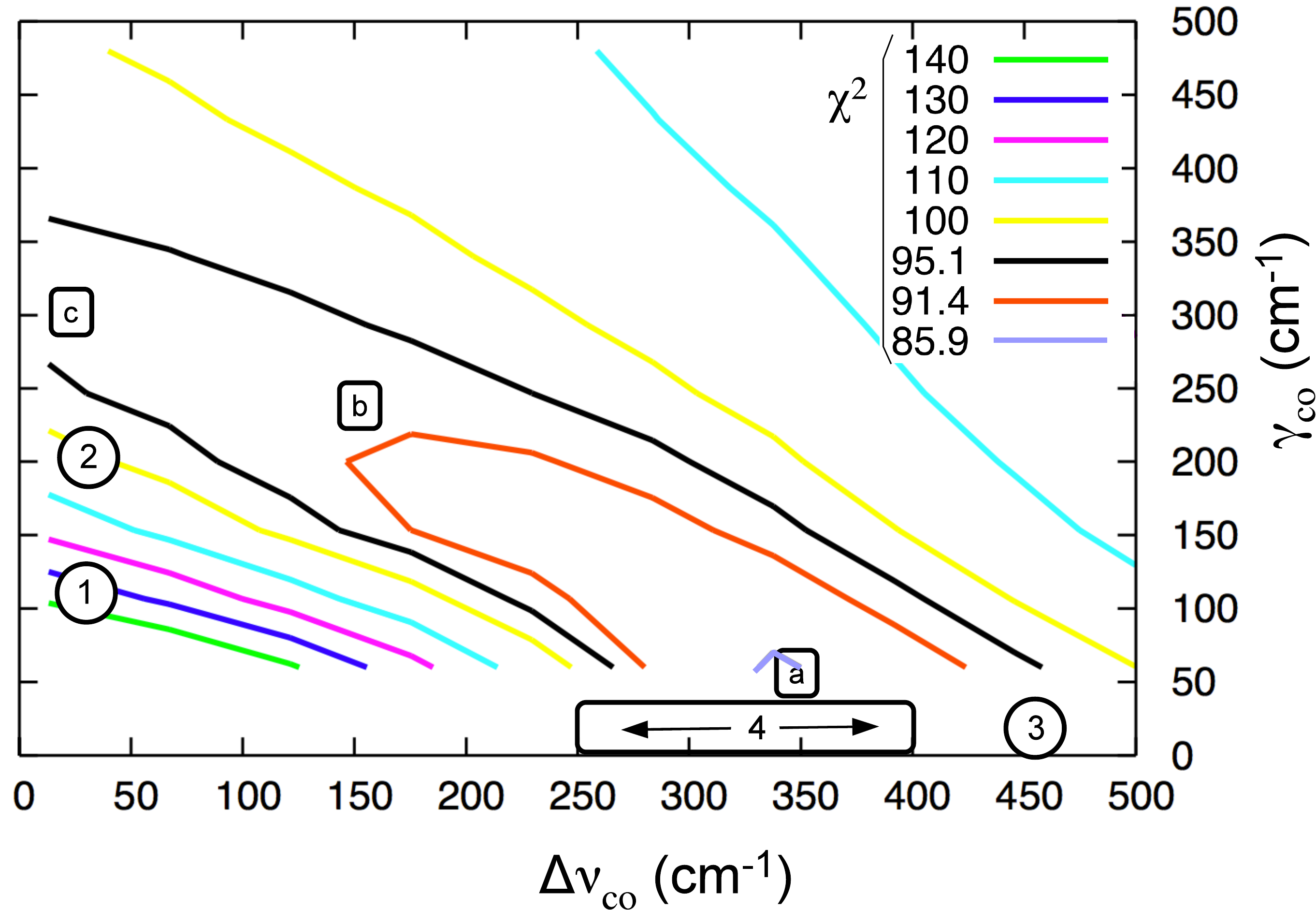}
  \caption{Map of the $\chi^2$ as a function of $\Delta \nu_{co}$ and $\gamma_{co}$, where $\chi^2$ is defined as $\sum\limits_{i=1}^{N} \frac{(I/F_\mathrm{model}-I/F_\mathrm{data})^2}{(\Delta I/F)^2}$ and $\Delta I/F_\mathrm{data}$,
  the uncertainty on data is set as $0.03 \times I/F_\mathrm{data}$ \citep[after ][]{Sromovsky2012}.
  The minimum value is around $\Delta \nu_{co}=\SI{350}{\per\cm}$ and $\gamma_{co}=\SI{50}{\per\cm}$ (label \fbox{a}).
  The value domain inside the $1-\sigma$ and $2-\sigma$ error are shown by the second contour ($\chi^2=91.4$ - orange) and third contour ($\chi^2=95.1$ - black).
  The $\chi^2$ obtained using reference values for the cut-off \citep[shown by the  encircled 1]{deBergh2012} is clearly outside the zone of the minimum $\chi^2$. On the other hand, cut-off parameters found in other works (\cbox{2} \cite{Hirtzig2013}, \cbox{3} \cite{Griffith2012} and \cbox{4} \cite{Bailey2011}) are more consistent with the zone of minimum $\chi^2$.
  Values ($\Delta \nu_{co},\gamma_{co}$) labelled \fbox{a}, \fbox{b} and \fbox{c} are used in the following as representative parameters for the region of minimum $\chi^2$.}
  \label{chi2map1}
\end{figure}

Our results show that we are able to find a common vertical structure whatever the profile cut-off parameters. The results obtained with 72 sets of ($\Delta \nu_{co}$, $\gamma_{co}$) yield $F_h = \num{2.2805 (888)}$, $X_2 = \SI{98.001 (1519)}{km}$, $X_1 = \SI{27.236 (73)}{km}$ and $X_3 = \SI{86.632 (853)}{km}$, where the uncertainties are given to $1-\sigma$.
The best profile, with the lowest $\chi^2$, is displayed in Figure~\ref{three_scale_heights} (curve labelled \textit{analytical function}). The uncertainty on $X_2$ is extremely small because this parameter is related to the part of the atmosphere between 225 and \SI{350}{km}, in the middle of the altitude range, where the radiance factor follows a scale height.
This vertical profile is quite similar to those reported ($H \simeq 60$ to \SI{69}{km} between 140 and \SI{176}{km}  altitude and $H \simeq  \SI{45}{km}$ between 176 and \SI{278}{km}) by \cite{Vinatier2010} in far infrared (7.04 to \SI{16.66}{\um}), at the same latitude (\ang{43.5}N and \ang{46.5}N) and about the same period of the year (flyby T16, July 2006).

\begin{figure*}[!ht]
\includegraphics[width=\textwidth]{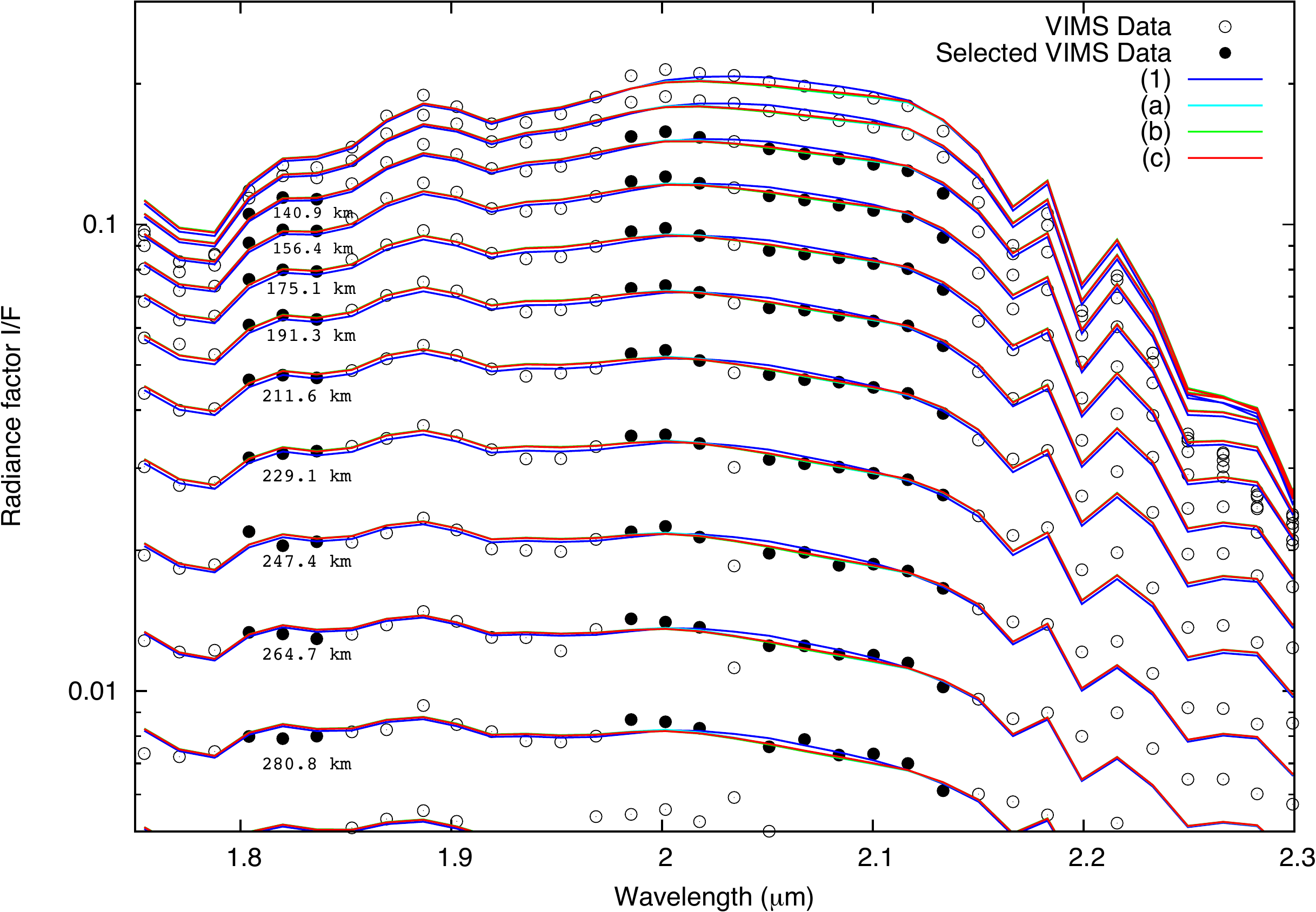}
\caption{As in Figure~\ref{spectra_fit0} except that model results are now
overplot to data. Only the data shown with filled circles are actually used in our analysis and the retrievals. The spectra we used are labelled with their altitudes, from 140 to \SI{280}{km}.
The blue curve (labelled \cbox{1}) corresponds to the results for the reference cut-off parameters and the curves labelled (\cbox{a}), (\cbox{b}) and (\cbox{c}) respectively correspond to cut-off parameters $\Delta \nu_{co} = \SI{350}{\per\cm}$ and $\gamma_{co} = \SI{60}{\per\cm}$, $\Delta \nu_{co} = \SI{150}{\per\cm}$ and
$\gamma_{co} = \SI{240}{\per\cm}$ and $\Delta \nu_{co} = \SI{13}{\per\cm}$ and  $\gamma_{co} = \SI{360}{\per\cm}$ taken in the zone of minimum $\chi^2$ (see Figure~\ref{chi2map1}). The curves labelled (\cbox{a}), (\cbox{b}) and (\cbox{c}) are extremely closed from each other and cannot be easily distinguished. Only data shown with a filled circle are used for the retrieval.}
  \label{spectra_fit1}
\end{figure*}

As for previous works \citep{Bailey2011, Griffith2012, Hirtzig2013} we find that the \SI{2}{\um} window is best fit with a cut-off applied further from the line center than the reference cut-off (Figure~\ref{chi2map1}).
The best results are obtained for values of $\Delta \nu_{co}$ and $\gamma_{co}$ along the line defined by the points ($\Delta \nu_{co} = \SI{50}{\per\cm}$, $\gamma_{co} = \SI{350}{\per\cm}$) and ($\Delta \nu_{co} = \SI{350}{\per\cm}$, $\gamma_{co} = \SI{50}{\per\cm}$).
In the following, we will refer to the cut-off parameters along this line as \textit{extended cut-off parameters}. There, the reduced $\chi^2$ values are around 0.85 (with the number of data minus parameters equal to $N-P=103$). For the reference value $\Delta \nu_{co} = \SI{26}{\per\cm}$ and $\gamma_{co} = \SI{120}{\per\cm}$, the reduced $\chi^2$ is significantly larger, around \num{1.3}. The exact values of these $\chi^2$ and reduced $\chi^2$ depends upon our knowledge of data uncertainties which is not well defined \citep{Sromovsky2012}.
The comparison between the data and the model for selected values of $\Delta \nu_{co}$ and $\gamma_{co}$ shows that the shape of the window is flatter and much better modelled with extended cut-off relatively to the reference choice (Figure~\ref{spectra_fit1}). The parameters of the cut-off published previously are shown on the $\chi^2$ map. They all show the same trend : the \SI{2}{\um}-window is best fit with an extended cut-off, and there are several ways to extend the cut-off.

\subsection{Absorption in the center the \SI{2}{\um} window}

As mentioned previously, there is a marked absorption in the centre of the \SI{2}{\um} window below \SI{300}{km} (Figure~\ref{spectra_fit0}). To perform our analysis, we exclude the channel (\#167) which potentially produces spurious values. Without this sharp peak, it still remains a broad and smooth absorption feature which is not accounted for by our model.
Our scope in this part is to check the impact of this feature on the model results. In particular, we want to know if adding an absorption can improve the results and make other choices of $\Delta \nu_{co}$ and $\gamma_{co}$ acceptable. The nature of this absorption feature is not known, and it may be produced by an absorption due to haze optical properties or it could be due to a gaseous species which may condense at the troposphere. From these hypotheses, we define three cases to include an absorption (Table~\ref{tab:abs_peak}).

\begin{table*}[!ht]
\caption{Different natures for the absorption peak}
\label{tab:abs_peak}
\begin{tabular}{l l l}
 \toprule
 Case  & Nature of the peak & Remark \\
 \midrule
 \#1  & Haze & Down to \SI{80}{km} / No absorption below \\
 \#2  & Gas with c$^\mathrm{ste}$ mixing ratio  & All the column \\
 \#3  & Condensible gas & Constant down to 60 km / No absorption below \\
 \#4  & Haze and mist & All the column \\
 \#5  & \ce{C2H6} GCM & From \cite{Rannou2006} \\
 \#6  & \ce{CH4} & From \cite{Niemann2010} \\
 \bottomrule
 \end{tabular}
\end{table*}

For the first case (\#1), we assume an absorption due to the haze layer by adding a Gaussian peak in aerosol absorption. We then add, for instance,  $\Delta \sigma_{abs}(\lambda)$ to the absorption cross-section.
It produces a corresponding increase in extinction ($\sigma'_{ext}(\lambda)=[\sigma_{abs}(\lambda)+\Delta \sigma_{abs}(\lambda)] +\sigma_{sca}(\lambda) =\sigma_{ext}(\lambda)+\Delta \sigma_{abs}(\lambda)$).
The scattering properties are not modified, and to conserve the same scattering properties, the single scattering albedo must be modified as following: $\varpi'(\lambda)= \varpi(\lambda) \times \sigma_{ext}(\lambda)/ \sigma'_{ext}(\lambda)$.
To mimic an absorption peak due to aerosols, we use $\Delta \sigma_{abs}(\lambda) = \sigma_0 \times f(\lambda)$, where $\sigma_0$ is a reference cross-section and $f(\lambda)$ is the Gaussian form, defined as $f(\lambda) = A \exp(-(\lambda-\lambda_0)^2/2\Sigma^2)$, where the amplitude $A$, the peak wavelength $\lambda_0$ and standard deviation $\Sigma$ can be left as free parameters.

\begin{figure}[!ht]
  \includegraphics[width=.45\textwidth]{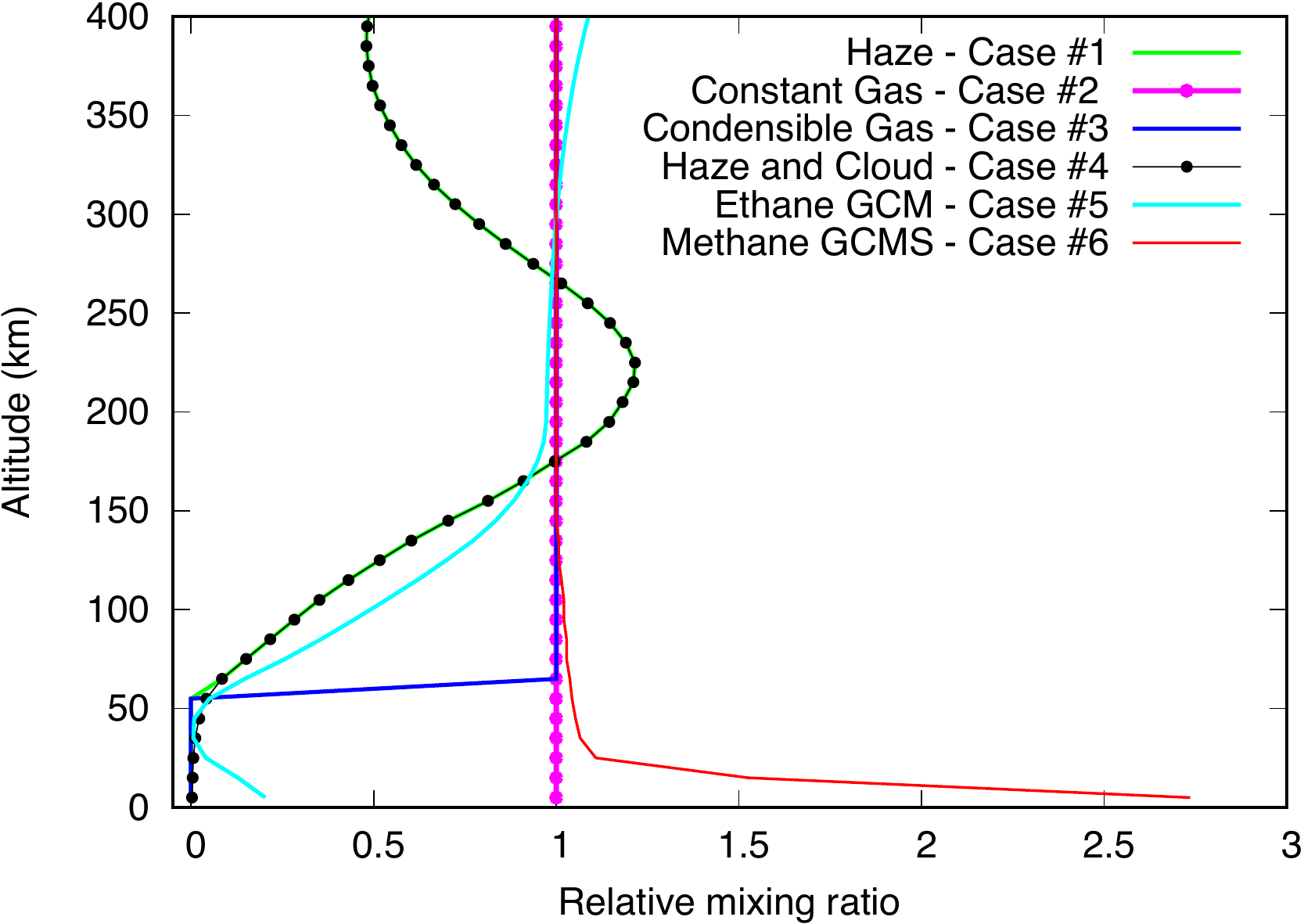}
  \caption{Vertical profiles of the absorption peak fed into the model for the various cases studied in this work. Profiles are given as relative mixing ratio (only the shapes of the profiles matter) and the actual absorption is further scaled by a free parameter.}
  \label{gas_profiles}
\end{figure}

In the second case (\#2), we assume an absorption due to a gas with a constant mixing ratio. In this case, we add a contribution to the gaseous absorption by scaling it on the absorption per molecule computed for the methane. Notably, in windows, methane absorption is a smooth flat continuum. We then add a contribution $\Delta\tau_g(\lambda,i)$ to the gas opacity $\tau_{g}(\lambda,i)$ that is written as follow : $\Delta \tau_{g}(\lambda,i) = f(\lambda) n_g(z) b(z,\lambda,i)$ where $z$ and $i$ are, respectively, the altitude and the index of the terms for the correlated-k calculation ($i=1,N$), $f(\lambda)$ is the form function, defined as previously, for the absorption peak, $n_g(z)$ is the total molecular concentration and $b(z,\lambda,i)$ the absorption coefficient of the methane used as reference for this putative gas.
Case \#3 accounts for the absorption peak  similar to the second way, except that the form function $f(\lambda)$ falls to zero below \SI{60}{km} in order to mimic the effect of a sharp condensation in the lower stratosphere (Figure~\ref{gas_profiles}).

\begin{figure*}[!ht]
  \includegraphics[width=\textwidth]{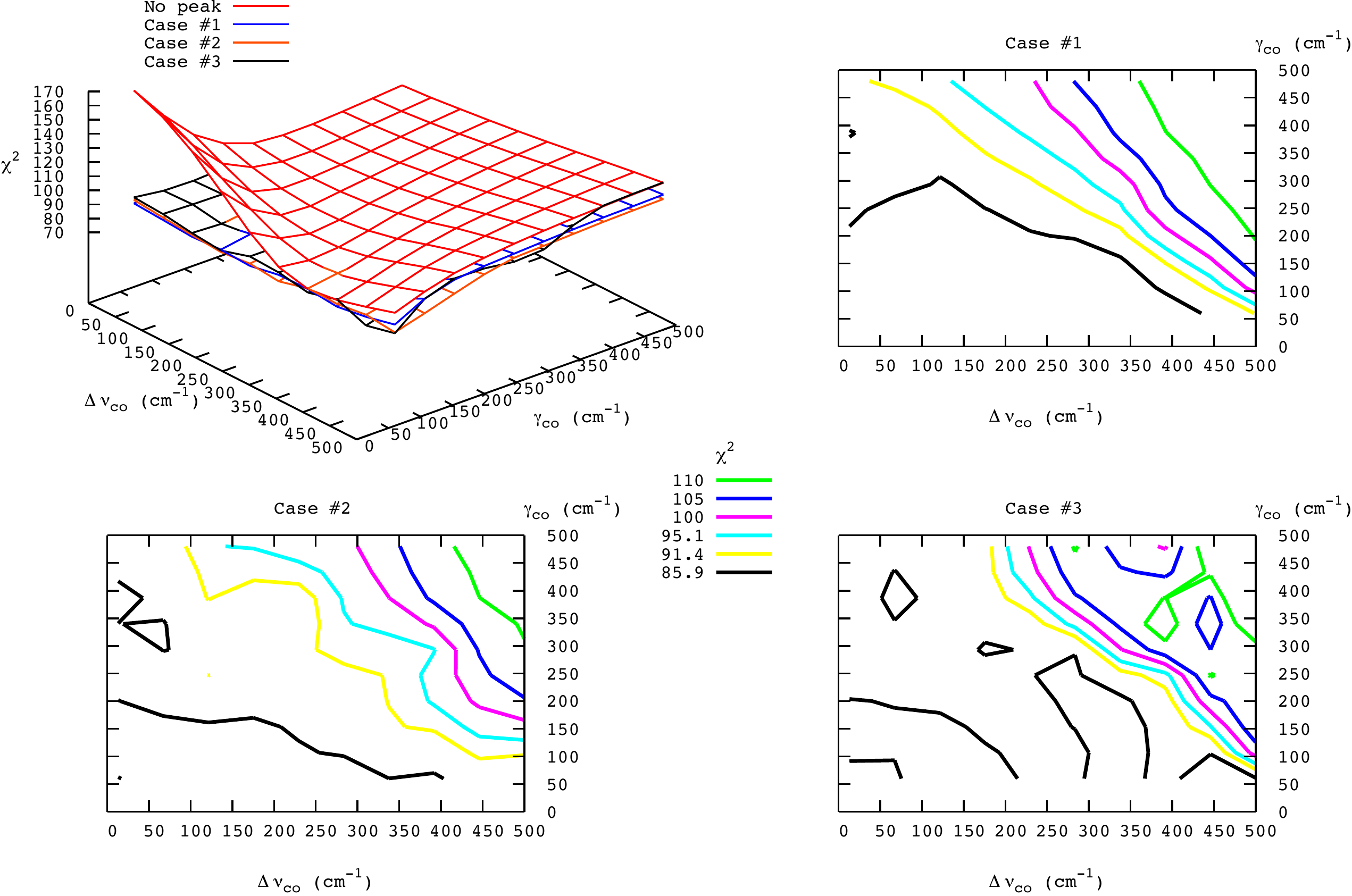}
  \caption{ (At top and left) Surfaces of $\chi^2$ as a function of $\Delta \nu_{co}$ and $\gamma_{co}$. The $\chi^2$ surface labelled \textit{No peak} corresponds to the case already shown in Figure~\ref{chi2map1}.
The other surfaces show the $\chi^2$ obtained with an absorption feature in the methane window. These maps show that the model results can be significantly improved if a supplementary absorption is included (case \#1, \#2 and \#3). For the case of the reference cut-off values ($\Delta \nu_{co} = \SI{26}{\per\cm}$, $\gamma_{co} = \SI{120}{\per\cm}$), the match between data and model is the same as for the extended cut-off.
The three cases give similar values of $\chi^2$, preventing us from preferring the fit of any one case.
(At top and right) Same as Figure~\ref{chi2map1}, and with the same $\chi^2$ levels, but for the case \#1 (the color code is displayed once for the three maps in the middle of the figure).
The two levels $\chi^2 = 91.4$ (yellow) and $\chi^2 = 95.1$ (cyan) corresponds to the $1-\sigma$ and $2-\sigma$ error level. With these new simulations, the region where the fits fall below the 1- or $2-\sigma$ levels is much broader than in the case \textit{No peak} and now includes the reference cut-off values.
(At bottom, left and right) Same as Figure~\ref{chi2map1}, for case \#2 and \#3.}
  \label{chi2_3Dmaps}
\end{figure*}

\begin{figure*}[!ht]
  \includegraphics[width=\textwidth]{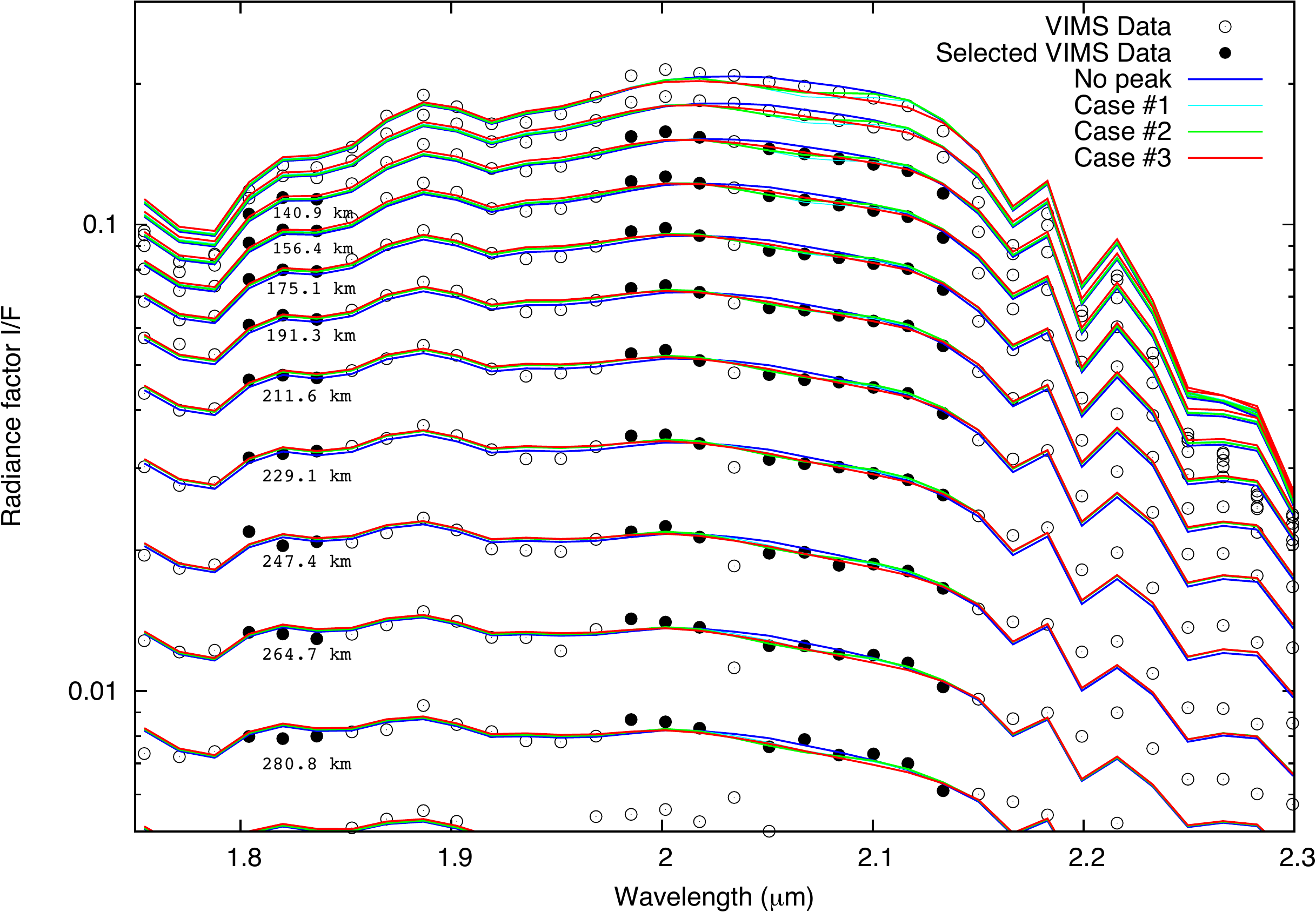}
  \caption{ As in Figure~\ref{spectra_fit1} except that we now overplot the results obtained for the reference cut-off values ($\Delta \nu_{co} = \SI{25}{\per\cm}$, $\gamma_{co} = \SI{120}{\per\cm}$), without peak (\textit{No peak}) and with an absorption peak for case \#1, case \#2 and case \#3. Only the data shown with filled circles are actually used in our analysis and the retrievals. The spectra we used are labelled with their altitudes, from 140 km to \SI{280}{km}.}
  \label{spectra_fit2}
\end{figure*}

For all cases, we perform a new set of analysis to improve the fits. This time, for each values of $\Delta \nu_{co}$ and $\gamma_{co}$, we used the corresponding set of scale heights found previously for the haze, and no longer considered as free parameters. The amount of haze $F_H$ is still considered as a free parameter as well as the three parameters of the Gaussian shape, $A$, $\lambda_0$ and $\Sigma$.
We use again a Levenberg-Marquardt routine to find the best solution, and we first consider the $\chi^2$ map for the three cases. For the three cases, we obtain similar results : if we add an absorption peak  we are now able to obtain fits that have the same statistical significance (same values of $\chi^2$) whatever the values $\Delta \nu_{co}$ and $\gamma_{co}$ (Figure~\ref{chi2_3Dmaps}).
This clearly means that any choice of parameters for the far wing cut-off can produce a valuable fit provided that an absorption feature is included. The $\chi^2$ maps indicate that fits are slightly better, although not in a statistically significant way, for cut-off with small values of $\Delta \nu_{co}$ and $\gamma_{co}$, as for instance the reference values of \cite{deBergh2012}. This can also be seen on the fits of $I/F$ spectra which are now very similar for any values of the parameters (Figure~\ref{spectra_fit2}).
Consequently, we conclude that the apparent discrepancy between the parameters of the far wing cut-off in different windows may be due to this absorption feature inside the \SI{2}{\um} window.

\begin{figure*}[!ht]
  \includegraphics[width=\textwidth]{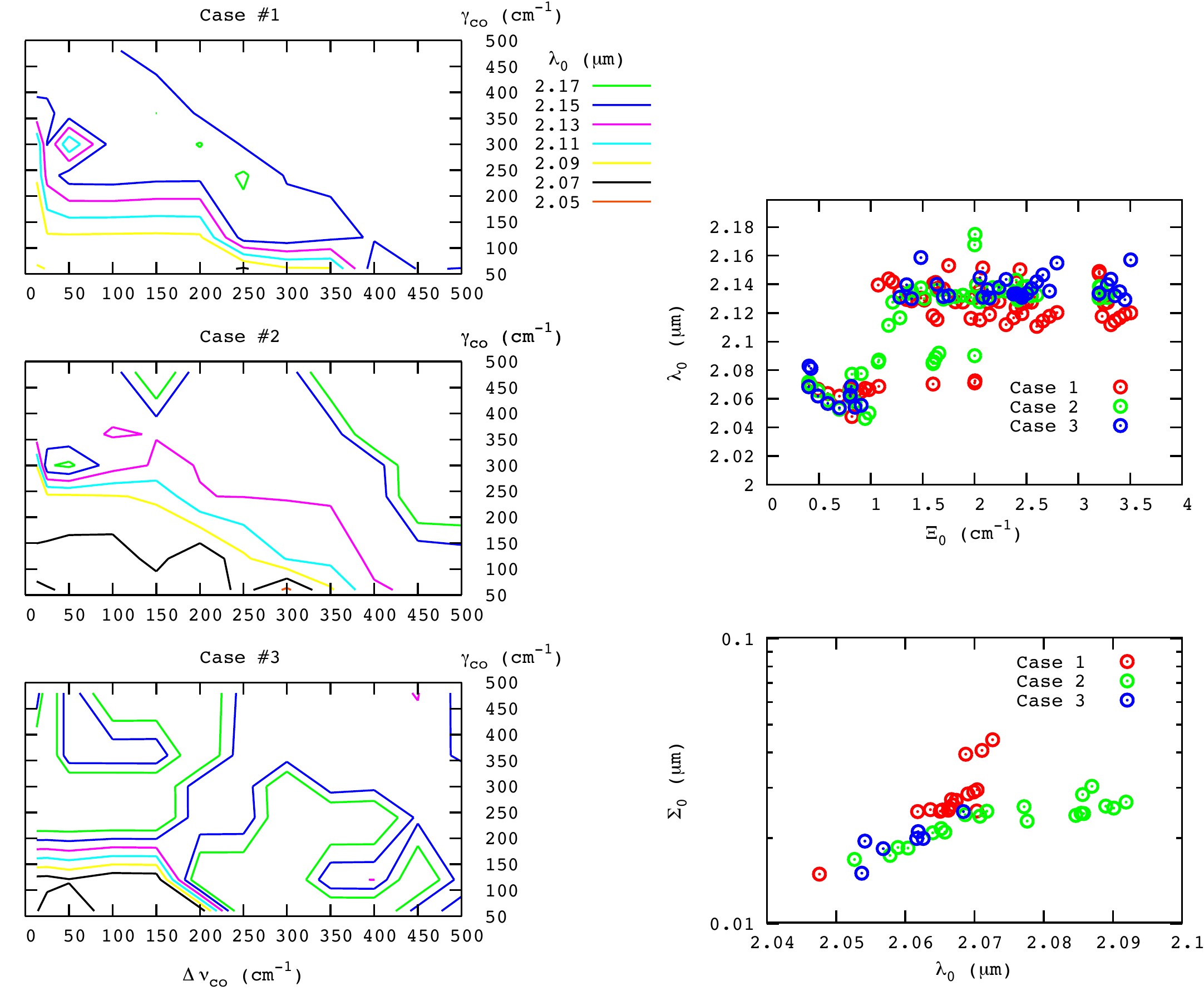}
  \caption{At left : Maps of the  absorption peak wavelength ($\lambda_0$) as a function of the $\Delta \nu_{co}$ and $\gamma_{co}$ for the three cases of absorption (case \#1, \#2 and \#3 from top to bottom ). The color code is the same for the three graphs and is shown for the case \#1 only.
  We can observe two distinct regions, with a peak between 2.05 and \SI{2.08}{\um} at small values of $\Delta \nu_{co}$ and $\gamma_{co}$ and another region at larger values of $\Delta \nu_{co}$ and $\gamma_{co}$ with $\lambda_0$ larger than \SI{2.10}{\um}. The transition between the two regions appears quite sharp.
  At right and top, the wavelength of the absorption peak $\lambda_0$ as a function of $\Xi_{co}$, defined as a metric in the  $\Delta \nu_{co}$ and $\gamma_{co}$ space as $\Xi_{co}=\sqrt{(\Delta \nu_{co}/150)^2 +(\gamma_{co}/350)^2}$.
  This graph clearly shows two populations of results, with  $\lambda_0$ smaller than \SI{2.1}{\um} for small values of $\Xi_{co}$ (that is for cut-off applied close to the core of the lines) and another population with  $\lambda_0$ beyond \SI{2.1}{\um} for extended cut-off. At right and bottom, the standard deviation $\Sigma_0$ of the peak as a function of the peak wavelength $\lambda_0$ (only for values smaller than \SI{2.1}{\um}) with the same color code as above.}
  \label{parameter_graphs}
\end{figure*}

Although the $\chi^2$ maps appear smooth, we can see a dichotomy in the parameters of the Gaussian absorption peaks which produces the results (Figure~\ref{parameter_graphs}). This occurs whatever the source of the absorption peak. Where the fit was the worse, at low values of $\Delta \nu_{co}$ and $\gamma_{co}$, we find that an absorption peak around $\lambda_0 \simeq 2.07$, with a dispersion around \SI{\pm 0.02}{\um} significantly improves the fit.
This occurs for parameters $\Delta \nu_{co}$ smaller than about \SI{200}{\per\cm} and $\gamma_{co}$ smaller than about \SI{150}{\per\cm}. For larger values of $\Delta \nu_{co}$ and $\gamma_{co}$, the best fit is found with a peak beyond \SI{2.10}{\per\cm} and this peak would reinforce the already existing absorption of the window side at long wavelengths (Figure~\ref{parameter_graphs}).
We excluded the range of wavelength beyond \SI{2.10}{\um} that we attribute to ethane, and the supplementary absorption may possibly be a residual absorption by the far side of ethane band around \SI{2.3}{\um}. It is important however to note that results where the peak is located around \SI{2.07}{\um} are significantly improved (in a statistical point of view) while results with a peak beyond \SI{2.10}{\um} are only marginally improved, as it can be concluded from the $\chi^2$ values.
Notably, the new $\chi^2$ values, obtained assuming an absorption peak, are very similar for the three cases, suggesting that they are equivalent despite corresponding to different ways to include the absorption.

\subsection{Absorption peak at \SI{\sim 2.07}{\um}}

We now focus on cases with a peak at $\lambda_0 < \SI{2.10}{\um}$, which corresponds to an absorption in the centre of the window. It should be noted that for all these solutions, the amount of haze is the same within a relative interval of \SI{\pm 6}{\percent} around an average value $F_H\simeq 2.17$. The root-mean square of the Gaussian absorption is around \SI{0.02}{\um}, but with $\gamma_{co}$ around 0.012 for $\lambda_0$ around \SI{2.05}{\um} and around 0.025 for $\lambda_0$ around \SI{2.09}{\um}.

It is not possible to compare directly the amplitude of the Gaussian function between different cases because the absorption is added to different components and with different rules. Rather, we have to compare impact of these additional absorptions on the atmosphere's extinction properties. We then consider the extinction coefficients of the haze and the gas at 4 different altitudes in the atmosphere for the three cases of absorption peak along with the opacity of the haze and gas in a model without an additional peak. We are especially interested by the absorption peak that is needed with the reference cut-off values $\Delta \nu_{co} = \SI{26}{\per\cm}$ and $\gamma_{co} = \SI{120}{\per\cm}$. Then, if we could also use it at \SI{2.0}{\um}, all Titan's spectra could be fitted with gaseous absorption treated in a consistent manner.

\begin{figure}[!ht]
  \includegraphics[width=.45\textwidth]{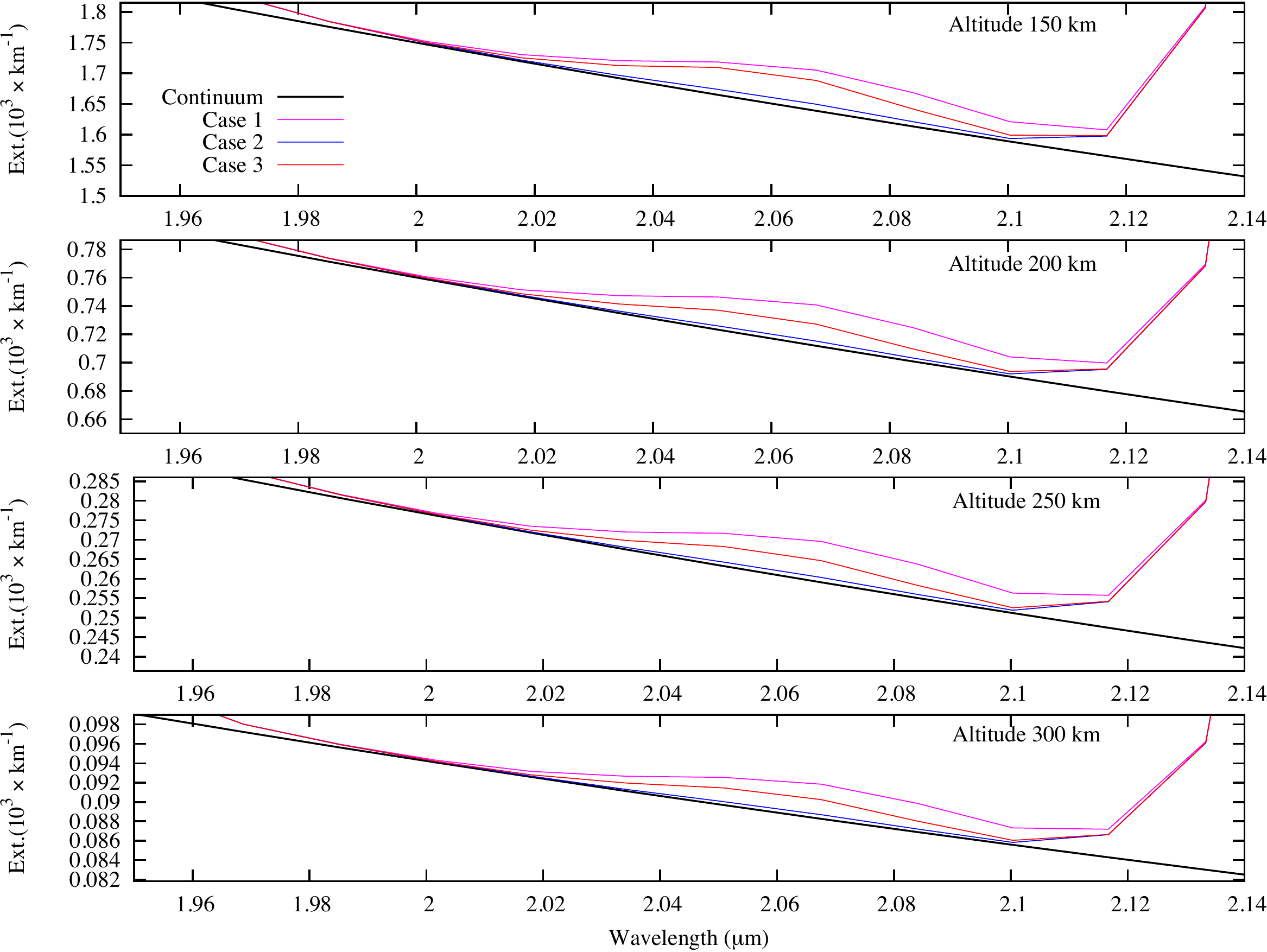}
  \caption{Extinction of the atmosphere at four altitudes for the three cases (\#1, \#2 and \#3), with the reference cut-off values ($\Delta \nu_{co} = \SI{25}{\per\cm}$, $\gamma_{co} = \SI{120}{\per\cm}$) along with the aerosols continuum.
  Although the three cases studied in this work gives comparable results for the scattered light at the limb, the corresponding extinctions appears very different and have more or less prominent absorption features.}
  \label{ext-peaks1}
\end{figure}

The prominence of the peak depends on the case we use to account for it. To obtain the same absorption feature in the scattered intensity we need an increasingly stronger peak if it is due to a non-condensible gas (case \#2), to a condensible gas (case \#3) or, finally, to the aerosol haze (case \#1) (Figure~\ref{ext-peaks1}).
To understand this, we must note first that, in our model, the decrease in intensities in the \SI{2}{\um} window is essentially produced by the influence of the multiple scattering due to the atmospheric column and not to the local optical properties of the atmosphere. This can be tested by setting the multiple scattering ratio $\rho_{ms}(z)$ to 1, without changing the atmosphere properties.
Then, the intensity obviously decreases and the absorption feature around \SI{\sim 2.07}{\um} completely disappears. This indicates that the absorption signature essentially comes from all layers through the multiple scattering and actually involves the entire atmospheric column.

In light of the these results, we must consider how the absorption mechanisms of our three cases are consistent with sourcing from the entire column. For instance, the cases \#2 and \#3 differs only because the absorption feature exists respectively along all the atmosphere column or only above \SI{60}{km}.
As a consequence, the absorption must be more marked in the case \#3 than in \#2 to compensate for the smaller column. If the extinction peak is due to aerosols (case \#1), then it has to be even stronger than for the gaseous cases. This is because the aerosol layer, in our model, goes down to \SI{80}{km} and has a different vertical profile than methane (Figure~\ref{gas_profiles}).
Below \SI{80}{km}, haze is replaced by a mist which does not bear any absorption signature. Therefore, the way the absorption peak is attributed (to haze or to gaseous properties) and the vertical distribution of the component which bears the absorption feature controls the depth of the absorbing layer.

To better understand the effect of the column depth on the absorption feature, we added three supplementary test cases for the vertical profile of the absorption feature. They represent variations around the previous cases. We only use these new profiles with the reference cut-off values $\Delta \nu_{co} = \SI{26}{\per\cm}$ and $\gamma_{co} = \SI{120}{\per\cm}$.
We tested the case with an absorption feature borned by both the stratospheric haze above \SI{80}{km} and mist below \SI{80}{km} (case \#4), the case of an absorption having a profile extracted from the 2D-IPSL Global Climate Model (IPSL GCM) for ethane (case \#5) and the case of an absorption having a methane-like profile (case \#6).
We introduce this last case to perform a test on the optical depth of the column because the methane profile sharply increases below \SI{40}{km}. But, since we are studying an absorption inside the methane window, we do not expect that this case may correspond to the reality.
These profiles are given in Figure~\ref{gas_profiles}, along with the three former profiles. We generally find that a smaller peak is needed with a gaseous absorption than with a haze absorption (not shown here).
We also find that a smaller absorption is needed when the vertical profile of absorption, whether it is due to haze or gas, goes deeper in the atmosphere. On the other hand, all these cases give statistically (in term of $\chi^2$) the same fit of the scattered intensity. We conclude that observations of the scattering intensity at the limb are not sufficient and that we need to consider other observations in order to go further.

\begin{table*}[!ht]
\caption{Best model parameters$^{\dag}$ for the reference cut-off$^{\ddag}$}
\label{tab:best_model}
\begin{tabular}{l l l l l}
\toprule
Case & $\lambda_0$ & $\Sigma$ & $A$ & $F_H$ \\
\midrule
 \#1 (Haze)  &   2.068$_{-0.023}^{+0.023}$&   2.640$_{-0.075}^{+0.083}\times$ 10$^{-2}$&   3.759$_{-0.053}^{+0.055}\times$ 10$^{-2}$&   2.115$_{-0.014}^{+0.014}$ \\
 \#2  (Gas c$^\mathrm{ste}$ mix. ratio)
                         &   2.064$_{-0.027}^{+0.027}$&   2.169$_{-0.078}^{+0.090}\times$ 10$^{-2}$&   4.143$_{-0.120}^{+0.130}$          &   2.114$_{-0.014}^{+0.014}$ \\
 \#3 (Condensible gas)   &   2.062$_{-0.023}^{+0.023}$&   2.074$_{-0.071}^{+0.081}\times$ 10$^{-2}$&   2.076$_{-0.063}^{+0.071}\times$ 10$^{+1}$ &   2.115$_{-0.014}^{+0.014}$ \\
 \#4 (Haze and mist)    &   2.068$_{-0.024}^{+0.024}$&   2.681$_{-0.077}^{+0.086}\times$ 10$^{-2}$&   3.042$_{-0.044}^{+0.047}\times$ 10$^{-2}$&   2.116$_{-0.014}^{+0.014}$ \\
 \#5 (\ce{C2H6} GCM)    &   2.063$_{-0.026}^{+0.026}$&   2.084$_{-0.078}^{+0.090}\times$ 10$^{-2}$&   1.569$_{-0.038}^{+0.041}\times$ 10$^{+1}$ &   2.113$_{-0.014}^{+0.014}$ \\
 \#6 (\ce{CH4})            &   2.068$_{-0.031}^{+0.031}$&   2.485$_{-0.096}^{+0.110}\times$ 10$^{-2}$&   1.758$_{-0.062}^{+0.072}$          &   2.115$_{-0.014}^{+0.014}$ \\
\bottomrule
\multicolumn{5}{l}{$\dag$ $\lambda_0$, $\Sigma$ and $A$ describe the peak properties while $F_H$ controls the amount of haze.} \\
\multicolumn{5}{l}{$\ddag$ Reference profile cut-off values are $\Delta \nu_{co} = \SI{26}{\per\cm}$ and $\gamma_{co} = \SI{120}{\per\cm}$ }\\
\end{tabular}
\end{table*}

The Table~\ref{tab:best_model} show the best peak parameters obtained for the reference cut-off, and for the six cases. Notably, the associated correlation matrices show that the four model parameters are well decorrelated since the highest non-diagonal term never exceed \num{3.5e-2}. The center and the width of the absorption peak, $\lambda_0$ and $\Sigma$, that we retrieve appear consistent whatever is the source of absorption (i.e the case).
Noteworthy, a more accurate value of $\lambda_0$ is approximatively \SI{2.065}{\um}, with a maximum uncertainty \num{\pm 0.03}. We do not expect the amplitude of the peak $A$ to be similar for the different cases because the absorption strength is defined relative to the absorption of the species which is assumed to bear the absorption.
This absorption differs from case to case. Therefore, only the resultant extinctions (as in Figure~\ref{ext-peaks1}), tangential opacities or transmissions through the atmosphere are relevant for comparison. This is the topic of the next section.

\begin{figure*}[!ht]
  \includegraphics[width=\textwidth]{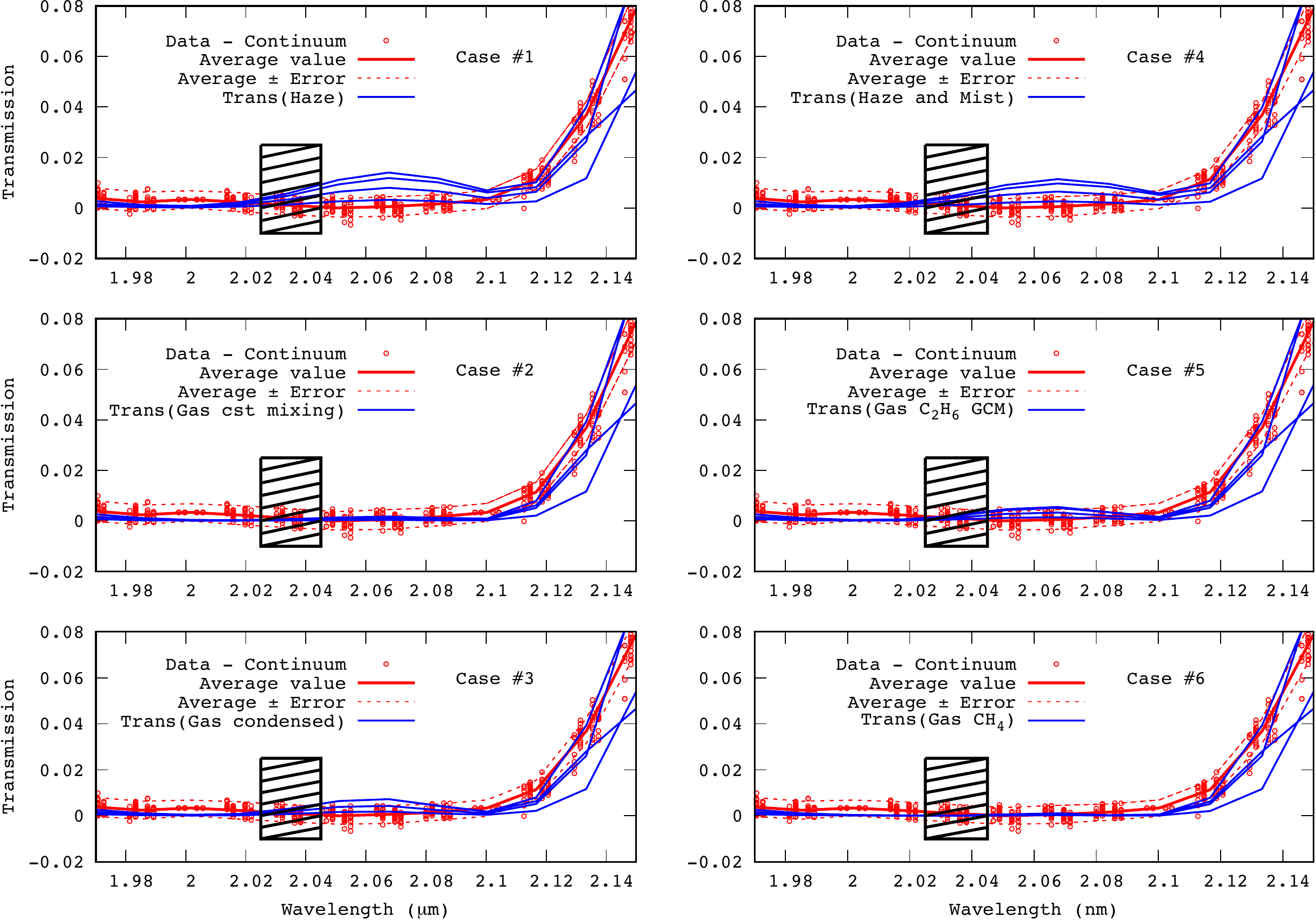}
  \caption{Relative difference ($(T - T_{s})/T$) between the transmission of the atmosphere ($T$) and the transmission of a synthetic haze continuum ($T_s$) as a function of the wavelength, in the range of altitude between 200 and \SI{300}{km}, obtained from stellar occultation observed by VIMS \citep[, red lines and symbols]{Maltagliati2015}.
  The rectangle with hatching represents a zone excluded from our analysis.  The calculations for the model (blue lines) are obtained for the reference cut-off values  $\Delta \nu_{co} = \SI{26}{\per\cm}$ and $\gamma_{co} = \SI{120}{\per\cm}$.
  The six figures correspond to the six cases of absorptions features, the three cases throughout this paper and the three additional cases (cf. Table~\ref{tab:abs_peak}). In these plots, an absorption inside the \SI{2}{\um} window should appear as a peak around \SI{2070}{nm}.
  The transmission computed by the model at four levels between 200 and \SI{300}{km} (blue lines) may contain such a peak depending on the cases while data does not contain this peak.}
  \label{occultation-peaks1}
\end{figure*}

\subsection{Occultations in the \SI{2.0}{\um} window}

To discriminate between the different solutions, we can compare the transmission at \SI{2}{\um} window through a tangential line of sight at the limb of Titan produced by the different solutions with real occultation transmission \citep[e.g., ][]{Bellucci2009, Maltagliati2015}.
None of the observations correspond to the latitude and the time period of the scattering observation. However, we also remark that the relative difference between the transmission through the atmosphere at different altitudes $T(z)$ and the transmission of a synthetic continuum, $T_s(z)$, are very similar for all the observations. We then decided to use the four profiles to get the average value of the $\Delta T(z)/T(z)=(T(z)-T_s(z))/T(z)$ and to evaluate the uncertainty on this value, including the intrisic error of the observation \citep{Maltagliati2015}.
For each observation, the synthetic continuum is assumed as a linear function computed from the value of transmission at \SI{2.0}{\um} (channel \#165) and at \SI{2.1}{\um} (channel \#171). In observations, the \SI{2}{\um} window does not seem to have a noticeable absorption feature at \SI{2.065}{\um}. We estimate that the potential dip in the transmission curve can not exceed 0.004 in term of transmission in the altitude range between \SI{200}{km} and \SI{300}{km}. This dip is barely perceptible in data while the various type of absorption profiles used in our model can yield more or less prominent peaks around \SI{2.065}{\um}, depending on the case.
For the case \#1, \#4, \#3 and \#5, we clearly see an absorption feature which exceeds the error level (Figure~\ref{occultation-peaks1}).
These cases can be discarded. The case \#6, only introduced for testing purpose, and the case \#2 do not produce absorption feature in tangential transmission. These cases are those with the largest integrated column and, consequently, those that need the smallest absorption peak. Therefore, only the case \#2 (a gas with constant mixing ratio) remains consistent with observations.
Ethane has a tiny absorption feature inside the methane window, as already noted by \cite{Maltagliati2015}, and could be a good candidate to explain this absorption feature.

The cross-sections of ethane as measured by \cite{Sharpe2004} give a set of spectroscopic structures between 2 and \SI{2.05}{\um}: the Q-branch for the pure vibrational transitions and, at the sides of the Q-branch, the P and R-branches for the ro-vibrational lines. These measurements were performed at \SI{5}{\degreeCelsius}, \SI{25}{\degreeCelsius} and \SI{50}{\degreeCelsius}, which is not representative of Titan's conditions.
However, this is the only source of information that we have and the reader should realize that ethane cross-sections at Titan's temperature could be slightly different.
The Q-branch matches quite well the sharp absorption in the channel \#167, although possibly shifted by few nanometers. The absorption produced by the P-branch, at the large wavelength side of the central Q-branch, may explain the actual shape of the window. However, only a complete model using spectroscopic data of ethane accounting for the actual Titan's conditions could allow us to check if an absorption in the middle of the \SI{2}{\um} could be explained by ethane.
Notably, we used an \emph{ad-hoc} Gaussian peak to mimic this absorption and we found its centre between 2.06 and \SI{2.07}{\um}. But the real shape of the peak due to the P-branch is asymmetrical with a peak slightly shifted toward short wavelengths and a longer tail toward large wavelengths. Finally, with spectroscopic data, we should also have an absorption feature at shorter wavelengths (due to the R-branch) around \SI{2.01}{\um}. With an accurate treatment of ethane spectroscopy, one must expect absorption patterns able to modify the opacity in the methane \SI{2}{\um} window.

It should be also noted that the GCMS onboard Huygens found a fairly constant mixing ratio for ethane with altitude, at the limit of the detection level \citep{Niemann2010}. However, the value appears quite uncertain in the low stratosphere and in the troposphere.
The upper value of ethane mixing ratio is given at $x_{\ce{C2H6}} = \num{e-5}$ and, despite error bars are given and higher values may be found in some layers, the final result is given as an upper value.
The GCMS also found a large quantity of several hydrocarbons and nitriles after Huygens landing. This revealed a wet surface with many condensing species, including ethane, in thermodynamical interaction with the atmosphere. We then expect the ethane mixing ratio to be different from 0, as assumed in the case \#3 (condensible gas) or in simple models where condensible species are removed below a given altitude. This also contradicts somewhat the results of the 2D-IPSL GCM (input for case \#5), which simulated Titan's climate, cloud microphysics and interaction between condensible gas, droplets and aerosols.
This model runs under some assumptions and approximations which could be revisited because they modify the vertical profile of ethane in the GCM. For instance, condensed methane and ethane on droplets are assumed to behave independently from each other. In reality, these condensed species, along with others, yield complex mixtures with thermodynamical laws differing from the laws of pure species.  This would drastically modify the saturation vapor pressures above droplets and, more generally, it would provide more complex equilibrium conditions between condensed and gaseous phases.

The influence of the source of ethane at the surface is not included in the GCM, although it is now known that the surface at the Huygens landing site was rich in liquid ethane \citep{Niemann2010} and that polar lakes also contain a fraction of liquid ethane as well \citep[e.g., ][]{Brown2008}.
Such effects are not accounted for in the 2D-IPSL GCM and they could contribute in increasing the abundance of ethane in the troposphere especially near the sources at surface. We then assume that the best option to account for ethane in this present work is to assume a constant mixing ratio down to the surface, which corresponds to case \#2.

\section{Impact on the retrieval of the surface reflectivity}

One important impact of having a good description of the opacity in the \SI{2}{\um} is the ability to retrieve the surface reflectivity. Previous works based on photometry retrieved surface spectra which suggested that Titan's surface could be made of water ice \citep{Griffith2003, Hirtzig2005, Rodriguez2006, Soderblom2009}.
The DISR instrument onboard Cassini observations yielded an \emph{in situ} evaluation of the surface reflectivity spectrum \citep{Tomasko2005, SchroderKeller2008, Karkoschka2012,Karkoschka2016}.
The surface spectra has a red slope from the visible to about \SI{0.9}{\um}, then a blue slope up to about \SI{1.5}{\um}, and finally a marked signature at \SI{1.6}{\um}, very similar to the water ice absorption signature. This spectra could be explained with a model of surface assuming a water ice layer covered by a mixture of aerosols and liquid \citep{Rannou2016}.
This model also predicts the surface spectrum beyond \SI{1.6}{\um}, and suggests that the water ice should leave strong signatures. In the \SI{2}{\um} window, the reflectivity should be seen as a U-shaped surface albedo, in the \SI{2.8}{\um} window, the reflectivity should decrease continuously through the window and at \SI{5}{\um}, it should be relatively flat or slightly increasing.

The importance of describing the spectral behaviour of surface reflectivity inside each window, instead of discussing the window-averaged reflectivities, is that we would be able to constrain specific identifying signatures. This requires a fine knowledge of the opacities inside the window, or at least of its spectral behaviour.
The \SI{2.8}{\um} window is too poorly known to have a safe inversion because aerosol absorption has a strong variation across this window \citep{Rannou2010} and several gaseous signatures shape this window. Ethane absorbs between 2.60 and \SI{2.70}{\um} and beyond \SI{2.8}{\um}, as deuterated methane at \SI{2.75}{\um}, and possibly other gases further than \SI{2.8}{\um}.
As mentioned previously, no ethane linelist is yet available for this wavelength range and the aerosol properties are very uncertain. Finally, the \SI{5}{\um} window can be used to retrieve the surface albedo, but for water ice we expect a flat and featureless behaviour which will not give a strong clues.
These arguments demonstrate the importance of understanding the \SI{2}{\um} window: with better understood opacities, it can reveal a specific signature  more reliably inverted than the \SI{2.8}{\um} window.

\begin{figure*}[!ht]
  \includegraphics[width=\textwidth]{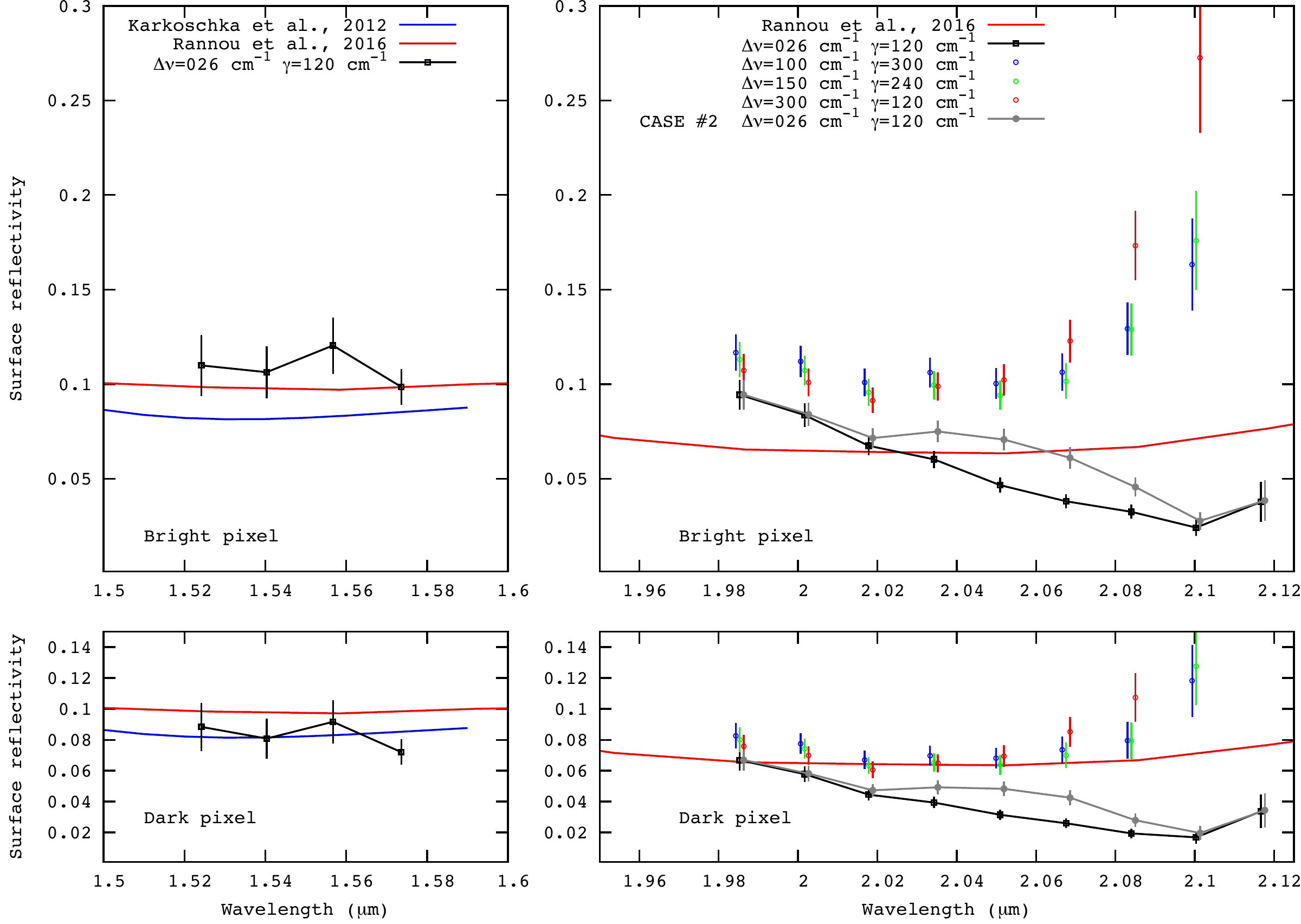}
  \caption{Surface reflectivities observed by DISR (blue line), modelled with a surface model (red line) and retrieved in this work for a bright and black pixel in the \SI{1.6}{\um} and \SI{2.0}{\um} windows. The error bars are given to $1-\sigma$.
  They essentially depends on the uncertainties on the radiance factor and on the uncertainty on the amount of haze. At \SI{1.6}{\um}, we used the reference cut-off values for the retrieval (dark line). At \SI{2.0}{\um}, we retrieved the surface reflectivity with different values of the cut-off parameters, the reference value and several extended values collected in region of low $\chi^2$ reported in Figure~\ref{chi2map1}, and no absorption peak.
  We also show the retrieved surface albedo obtained with the reference value of the cut-off parameters and with the absorption peak corresponding to the case \#2.}
  \label{Surface_reflectivity}
\end{figure*}

To retrieve the surface albedo, we use an observation taken during the flyby of Titan T71. This is the same image as in \cite{Rannou2016}, chosen because it has a moderate solar incidence (\ang{\sim 35}) and emergence (\ang{\sim 26}) angles and a fairly good spatial resolution (\SI{10}{km}) that allows to see quite homogeneous regions of dark and bright surfaces.
This characteristics are fine for the purpose of the present study. We use two pixels, one taken in the bright zone and the other one in the dark zone. We accounted for a wavelength shift of \SI{5.6}{nm} in the data recording of VIMS. In the model we strictly use the setup published by \cite{Doose2016} for the haze.
We then use their vertical profile as a reference for both the haze and the mist. The spectral properties of haze and the gaseous absorption are computed in the same way has described for this study at the limb. We study \SI{1.6}{\um} and \SI{2.0}{\um} windows. For a given spectrum, we first use the intensity in the methane band to set the amount of haze and mist, scaling the reference vertical profile, with the factor $F_H$. Once the value of $F_H$ is set, we used the complete spectrum to retrieve the surface reflectivities ($A_i$ with $i=1,N$) which remains the only free parameters.
The index $i$ runs over the number of channels which probes the surface. We use 13 different values : four values for the \SI{1.6}{\um} window, and nine values for the \SI{2}{\um} windows. These numbers of channels are essentially constrained by the spectral widths of these windows which give access to surface information.

We first retrieve the surface albedo without including absorption peak, but with different sets of cut-off parameters. Again we choose the reference values  ($\Delta \nu_{co} = \SI{26}{\per\cm}$, $\gamma_{co} = \SI{120}{\per\cm}$) and cut-off parameters chosen in the region of minimum values of $\chi^2$ (Figure~\ref{chi2map1}):
($\Delta \nu_{co} = \SI{100}{\per\cm}$, $\gamma_{co} = \SI{300}{\per\cm}$), ($\Delta \nu_{co} = \SI{150}{\per\cm}$, $\gamma_{co} = \SI{240}{\per\cm}$)
and ($\Delta \nu_{co} = \SI{300}{\per\cm}$, $\gamma_{co} = \SI{120}{\per\cm}$).
As a result, we find a large difference between the retrieval obtained with the reference cut-off values, which give a decrease of the surface albedo with the wavelength, and the retrieval with extended cut-off which gives U-shaped surface spectra (Figure~\ref{Surface_reflectivity}).
The surface reflectivity retrieved with the reference set is very similar to previous results by \cite{Negrao2006, Negrao2007} and \cite{Cours2010} for instance. Then, we now consider again the parameter set ($\Delta \nu_{co} = \SI{26}{\per\cm}$, $\gamma_{co} = \SI{120}{\per\cm}$) but, this time, with an absorption peak as in the case \#2.
We find that the retrieved surface albedo is flatter than in the reference case and do not produce a U-shape as the cases with extended cut-off. Adding a peak with the reference cut-off, as we do in this work, essentially modifies the surface albedo between 2.02 and \SI{2.10}{\um}.
Although the absorption feature is accounted for in a too simple way in this work, as discussed later, this demonstrates that the additional peak needed to explain the \SI{2}{\um} window at the limb may also have an impact on the retrieved surface albedo.

\begin{figure}[!ht]
  \includegraphics[width=.45\textwidth]{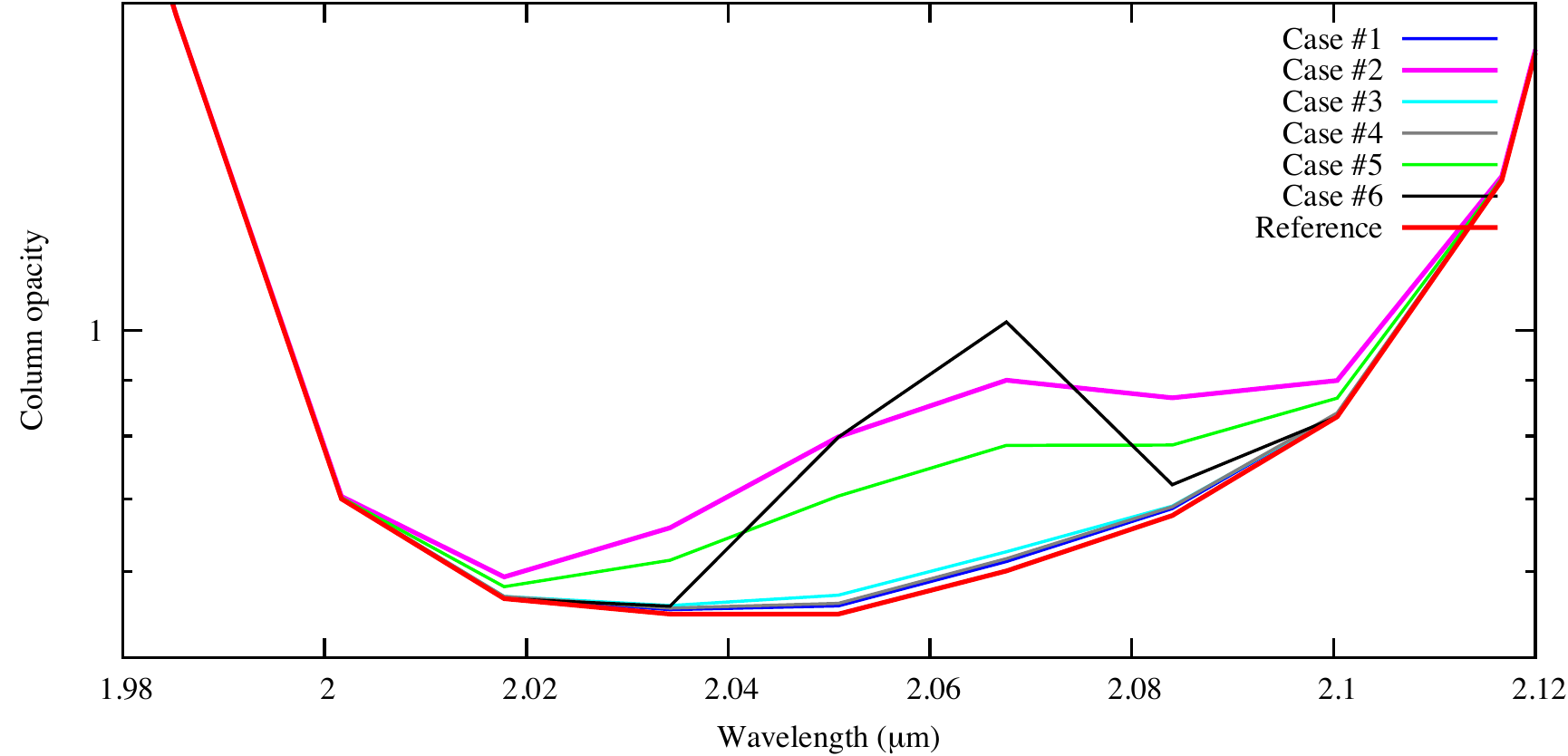}
  \caption{Total column opacity of the atmosphere for the simulation that yielded the results displayed in Figure~\ref{Surface_reflectivity}, for the reference cut-off values, and with the absorption cases \#1 to \#6.}
  \label{Integrated_columns}
\end{figure}

\section{Discussion and Conclusion}

There are still some sources of uncertainties in our analysis that must be considered and could have potentially an impact on the result.  We stress that this absorption peak, presumably gaseous in nature, is represented by a Gaussian function, with an identical form at all altitudes, which may differ from a real gaseous absorption peak for at least three reasons.
First, a gaseous peak essentially depends on the pressure level and on temperature which are not accounted for here. Secondly, if such an isolated absorption can produce a peak, its shape is probably not Gaussian.
Finally, we excluded some spectral regions from our study where ethane is known to contribute to the absorption, however this gas may participate to the continuum absorption beyond this exclusion regions and also may alter the results. We know that the retrieval of the reflectivity is very sensitive to the total column opacity.

Despite these limitations, our results demonstrate that the \SI{2}{\um} window can be fitted with the same line profile cut-off than other windows. For this, we need to include a supplementary peak of absorption at \SI{2.065 (27)}{\um} (result for case \#2 in Table~\ref{tab:best_model}).
This absorption feature, which appears clearly in the photometric observations at the limb, is treated in a \emph{ad hoc} manner with a Gaussian absorption. This mimics a gaseous absorption, since an absorption by particles (haze or cloud) is clearly excluded by our analysis. Such a supplementary absorption has a consequence on the retrieved surface albedo.
While the reference values for the line profile cut-off produces a monotonically decreasing reflectivity as already found in previous works, the absorption inside the \SI{2}{\um} window modifies significantly the shape of the surface albedo. The result is difficult to examine in detail because the new retrieved surface albedo depends on the way the absorption peak is accounted for, and our procedure does not ensure an exact retrieval of the surface reflectivity.
This result only shows that we have to expect a different surface spectrum at \SI{2}{\um} when more accurate data will be available to treat ethane absorption \citep{Viglaska2017}.

We also note that using line profile cut-offs with larger extension, as those giving the minimum $\chi^2$ in the initial test, may give satisfactory results.
These parameters allow us to fit the limb photometry and the occultation quite well. Moreover, the surface reflectivity that we can retrieve with these parameters are higher and have a U-shape that would be expected for a water ice signature. However, this case alone would not explain the tiny absorption in the centre of the \SI{2}{\um} window.
Some solution including these extended cut-off and a supplementary absorption to produce a dip in the spectra of scattered  light at the limb may also be considered.
Although we can not exclude this solution, we think more likely that the line profile cut-off are similar at all windows, which favor the solution with the reference cut-off values and a peak as described in this work.

This work strongly suggests that ethane is responsible for this absorption. Although we removed the spectral intervals where ethane has its most prominent absorption features, we noticed a small absorption feature in the ethane absorption pattern at a wavelength larger than \SI{2}{\um}, right in the centre of the window \citep{Maltagliati2015}. We found that to be consistent with both the observations of scattered light at the limb and the occultation, this absorption peak should be borne by an element following a constant mixing ratio, a characteristic of ethane \citep{Niemann2010}.

\section*{Acknowledgment:}
This work has been performed thank to the Agence Nationale de la Recherche (ANR Project \textit{APOSTIC} No. 11BS56002, France). The authors thank Dr S. MacKenzie for her helpful review and for significantly improve this paper.


\section*{References}
\bibliography{Biblio}

\end{document}